\documentclass{article}

\usepackage{rotating}
\usepackage{amsmath}

\input{epsf.tex}

\begin{document}

\title{Combustion waves in a model with chain branching reaction and their stability}

\author{V. V. GUBERNOV$^\ast$\dag\thanks{$^\ast$Corresponding
author. E-mail: vvg@spmlab.phys.msu.su}, H. S. SIDHU\ddag $~$and
G. N. MERCER\ddag\\\vspace{6pt}
\dag P.N. Lebedev Physical Institute of Russian Academy of Science,\\
Department of Theoretical Physics,\\ 53, Leninskii prospect,
Moscow 119991, Russian Federation\\~\\
\ddag School of Physical, Environmental, and Mathematical Sciences,\\
University of New South Wales at the Australian Defence Force
Academy,\\ Canberra, ACT 2600, Australia}

\maketitle

\begin{abstract}In this paper the travelling wave solutions in the adiabatic model
with two-step chain branching reaction mechanism are investigated
both numerically and analytically in the limit of equal
diffusivity of reactant, radicals and heat. The properties of
these solutions and their stability are investigated in detail.
The behaviour of combustion waves are demonstrated to have
similarities with the properties of nonadiabatic one-step
combustion waves in that there is a residual amount of fuel left
behind the travelling waves and the solutions can exhibit
extinction. The difference between the nonadiabatic one-step and
adiabatic two-step models is found in the behaviour of the
combustion waves near the extinction condition. It is shown that
the flame velocity drops down to zero and a standing combustion
wave is formed as the extinction condition is reached. Prospects
of further work are also discussed.
\end{abstract}



\section{Introduction}

Combustion waves have been studied for some time and are topic of
a relatively recent review \cite{Merzh}. They have been observed
in numerous experiments \cite{Merzh} and play an important role in
industrial processes, such as the production of advanced materials
using Self-propagating High-temperature Synthesis (SHS)
\cite{Makino}. The one-step irreversible reaction models have
contributed greatly to our understanding of these phenomena. In
these models it is assumed that the reaction is well modeled by a
single step of fuel ($F$) and oxidant ($O_2$) combining to produce
products ($P$) and heat. The generic kinetic schemes of models
with one-step chemistry are: $F\longrightarrow P + heat$ or $F +
O_2\longrightarrow 2P + heat$, where the temperature dependent
rate of the reaction is given by Arrhenius kinetics $k(T) =
e^{-T_a/T}$ and $T_a$ is the activation temperature. These models
have proven their usefulness since they are relatively simple and
allow analytical investigation using asymptotic methods in the
limit of infinitely large activation temperature \cite{Zeld,SIAM}.
The most important feature of one-step models is that they have
led to many useful and qualitatively correct predictions for
phenomena such as: ignition, extinction and stability of diffusion
flames; propagation and stability of premixed flames; flame balls
and their stability; structure and propagation of flame edges etc.

However, in the overwhelming majority of cases, the chemical
reactions in flames proceed according to a complex mechanism, that
involves a variety of different steps \cite{Zeld}. Moreover, for
many reactions, models with simple one-step kinetics may lead to
erroneous conclusions as noted in \cite{West}. In other words, if
we want to obtain a realistic description of the flame kinetics,
several different steps, each with its own intermediate chemical
species, have to be taken into account. Recent advances in
computational power have made it possible to study the flame
behaviour using full numerical solutions of the equations of
energy and mass transfer for all of the species involved with
detailed chemistry. Several numerical codes have been developed in
order to carry out these calculations, such as Sandia's CHEMKIN
code \cite{Warn} or the Flame Master code \cite{FM}. These
numerical algorithms have been successfully applied to analyze the
properties of both diffusion \cite{San06} and premixed flames
\cite{San00}. Although such investigations are useful in providing
some quantitative prediction for observed phenomena, there is
still a great deal of uncertainty about the reliability of these
complex models when applied to the prediction of generic behaviour
of flames such as stable combustion regimes, limits of the flame
extinction and particularly the onset of combustion phenomena such
as pulsating and cellular combustion, when the reactions begin to
change rapidly in space and/or time.

A logical compromise between the models with singe-step and
detailed multi-step kinetics have been found recently in using
reduced kinetic mechanisms. In a number of papers
\cite{San06,San00,Betch,Sesh94} (listing only few of them) the
detailed schemes of the hydrogen and methane oxidation, which
include dozens of intermediate reactions, were successfully
reduced to several steps. The remarkable feature of these type of
models is that they allow on one hand, analytical investigation
\cite{Betch,Sesh94} to be successfully undertaken, while on the
other hand they are able to produce excellent quantitative results
\cite{San06,San00} to accurately predict the main flame
characteristics for some specific reactions such as flame
propagation velocities, flame structures including the profiles of
temperature and reactants etc. These studies are very important
for applied problems, in which the properties of flames for
specific reaction and flame configuration are studied.

The importance of investigation of flames with complex kinetics is
well acknowledged in combustion literature and as noted in Clavin
et al. \cite{Clavin87} ``a crucial first step in such a study
consists in determining reduced schemes that capture the essence
of complete network''. A particulary promising candidate is a
two-step reaction mechanism. In this paper we only consider
combustion waves in premixed flames. For premixed flames, several
models involving two chemical steps have been considered
previously. These studies address the fundamental (generic)
properties of flames with multi-step reaction mechanisms rather
than consider some specific combustion reaction. The two-step
reactions can be split into two groups: reactions with competing
steps \cite{Zeld,Clavin87} and sequential steps
\cite{Zeld,Joulin85,Dold03,Dold04,Simon,Please03,ProA}. In the
first group of models three types of reactions are considered: (i)
simple competing scheme $\bf C_1$: $A\rightarrow P_1$,
$A\rightarrow P_2$; (ii) competing chain reaction scheme $\bf
C_2$: $A\rightarrow B$, $A + B \rightarrow P$; (iii) competing
chain branching scheme $\bf C_3$: $A + B \rightarrow 2B$, $A + B
\rightarrow P$, where $A$ is the reactant, $B$ the radical, $P$
the product. The second group of the reactions includes several
schemes: (i) non competitive reactions $\bf N_1$: \cite{Simon}
$A_1\rightarrow P_1$, $A_2 \rightarrow P_2$; (ii) sequential fuel
decomposition reaction $\bf N_2$: \cite{Zeld,Dold03,Please03}
$A\rightarrow B \rightarrow P$; (iii) chain branching reaction
$\bf N_3$: \cite{Zeld,Joulin85,Dold03,Dold04,ProA} $A + B
\rightarrow 2B$, $n B + M\rightarrow n P +M$, where $n=1$ in
\cite{Dold03,Dold04} and $n=2$ in \cite{Zeld,Joulin85} and $M$ is
a third body. These schemes were investigated either analytically
by using high activation energy asymptotic or numerically. As a
rule the asymptotic methods allow the investigation only in
certain distinguished limits when the inner equations arising in
the asymptotic procedure can be integrated analytically
\cite{Clavin87,Dold03,Dold04,Please03}. However, generally these
equations cannot be integrated analytically due to the increased
complexity of the two-step models in comparison to one-step
reaction models and the inner equations are solved numerically to
match them to the outer equation solutions
\cite{Sesh94,Clavin87,Joulin85}. In most cases the two-step
reaction models show more rich dynamical behaviour in comparison
to their single step counterparts. For given parameter values
there can exist combustion wave solution branches travelling with
unique speed, with two different speeds (C-shaped dependence of
speed on parameters), and with three different speeds (S-shaped
dependence of speed on parameters). The multiplicity of the flames
suggests that a more complex stability behaviour also exist.
However, the stability analysis of these flames are reported in
only a few papers \cite{Clavin87,Dold04} and has not been
investigated systematically. Since the behaviour of multi-step
models can differ from their one-step prototypes, we can expect a
number of new phenomena such as bistability (which cannot be
predicted by the one-step models) to be found as a result of such
an analysis \cite{Clavin87}.

In this paper we focus on the investigation of premixed combustion
waves in a model with non-competing two-step chain branching
reaction. As noted in \cite{Dold03} the chain branching reaction
is more realistic in describing real flames such as hydrocarbon
flames in comparison to other two-step reaction models. This model
was initially considered in \cite{Zeld} and then generalized in
\cite{Joulin85}. It is usually referred to as
Zeldovich-Li\~{n}\'{a}n model. In \cite{Joulin85} the model is
considered in a limiting case when the recombination step is fast
in comparison to heat conduction time. Firstly,the resulting
problem is investigated semi-analytically by using activation
energy asymptotic first which provides a set of equations for the
inner problem. These equations were then solved numerically. It
was shown that there exists a unique travelling wave solution for
given parameter values. In \cite{Sesh94} a reduced two-step chain
branching reaction for oxidation of hydrogen was studied. The
methodology used in this paper is similar to \cite{Joulin85} (i.e.
semi-analytical investigation using activation energy asymptotics
and numerical integration of the resulting equations). It was
shown that the flame has a unique burning velocity. The results
obtained in this paper are shown to be in a good agreement with
numerical experiments for the detailed kinetic mechanism of the
reaction, which indicates that the two-step reaction models might
have a capacity of quantitatively predicting the flame behaviour.
There are other papers devoted to investigation of the premixed
flames in this model and we refer the reader to \cite{Dold04} for
more detailed overview, however the stability of combustion waves
in the Zeldovich-Li\~{n}\'{a}n model has not been investigated. In
\cite{Dold03,Dold04} a simplified version of
Zeldovich-Li\~{n}\'{a}n model is introduced. In this model the
order of the recombination reaction is taken as $n=1$. In contrast
to \cite{Zeld} the model also includes the linear heat loss to the
surroundings. The model is studied in the limit of high activation
energy which allows asymptotic analysis to be carried out
successfully. The speed of the combustion wave is determined as a
function of the parameters of the problem. It is shown that in the
nonadiabatic case the dependence is C-shaped. For the adiabatic
case the expression derived in \cite{Dold04} suggests a unique
speed of the flame propagation. It is remarkable that in
\cite{Dold04} the stability analysis of combustion waves is
carried out for the case of the reactant Lewis number less than
one. The analysis predicts that the wave can lose stability due to
the cellular instabilities emerged in this case. In our previous
paper \cite{JMC06} we have investigated the properties of the
model introduced in \cite{Dold04} in the adiabatic case and in the
limit of equal diffusivity of the reactant, the radical and heat.
In contrast to \cite{Dold04} the activation energy in \cite{JMC06}
is taken to be an arbitrary number (not an infinite number) and we
used a different nondimensionalization. Our non dimensional
parameters are common for one-step models, which makes for an
easier comparison between the two models. In \cite{JMC06} the
properties of the travelling wave solutions are investigated in
detail by means of numerical simulation. It is demonstrated that
the speed of combustion wave as a function of parameters is single
valued. We have found that for finite activation energy there is a
residual amount of reactant left behind the travelling combustion
wave which is not used in the reaction. This makes the problem
similar to the nonadiabatic one-step premixed flames. The other
fact about the model considered in \cite{JMC06} which makes the
similarity between the adiabatic two-step reaction and the
nonadiabatic one-step system even stronger is that, at certain
parameter values the combustion wave exhibits the extinction.
However, it occurs in such manner that the speed of the wave drops
to zero. This mainly distinguishes the one- and two-step models,
since in the former case the wave can propagate for any parameter
values and even in the nonadiabatic case the speed of combustion
wave does not vanish for finite parameter values, whereas this is
not the case for the two-step models.

Extinction of the combustion wave is a critical property of the
model that also distinguished the two-step models form their
one-step counterparts. Therefore we have also examined this
property in detail in this paper. The other important question
which will be studied here is the stability of the combustion
wave. We will compare our results (where possible) with the
predictions of the asymptotic analysis of \cite{Dold04}. In
\cite{JMC06} the existence of multiple travelling wave train
solutions is reported for some parameter values. Our analysis
shows that besides wave trains, the solutions with various
structure can also exist such as pulses, fronts with step-like
structure etc. The properties and relevance of these solutions
lies beyond the scope of this paper and will be discussed in
future work. In this paper we will only consider the classical
travelling front solutions.

The rest of the paper is organized as follows. The nondimensional
model equations are introduced in section \ref{model}, and where
possible, we have referred the reader to \cite{JMC06} to avoid
repetition. In this section we also outline the formulation of the
problem to obtain the travelling front solution and determine the
condition for the  existence of the travelling waves. Section 3 is
devoted to the investigation of the travelling wave solution. In
the first part of section 3, we summarize the properties of this
solution which were obtained in \cite{JMC06}. We then determine
the extinction limit for the travelling front solution. We analyze
its behavior near the point of extinction using the matched
asymptotic expansion method and compare the predictions of
analytical and numerical approaches. We conclude section 3 by
undertaking the stability analysis of the travelling wave
solutions. Finally, in section 4 the concluding remarks are
presented with some discussion for future work.

\section{Model}\label{model}

We consider an adiabatic model (in one spatial dimension) that
includes two steps: autocatalytic chain branching $A+B \rightarrow
2B$ and recombination $B + M \rightarrow C+M$. Here $A$ is the
fuel, $B$ the radical, $C$ the product, and $M$ a third body.
Following \cite{Zeld,Dold03,Dold04} we assume that all the heat of
the reaction is released during the recombination stage and the
chain branching stage does not produce or consume any heat. As
noted in \cite{Zeld}, in this scheme the recombination stage
serves both as an inhibitor which terminates the chain branching
and an accelerant which produces heat. According to \cite{JMC06},
the equations governing this process can be written in
nondimensional form as
\begin{equation}\label{PDE}
\begin{array}{ccl}
u_t&=& u_{xx} + r w,\\
v_t&=&\tau_A v_{xx} - \beta vw e^{-1/u},\\
w_t&=&\tau_B w_{xx} + \beta vwe^{-1/u} - r \beta w,
\end{array}
\end{equation}
where $u$, $v$ and $w$ are the non-dimensional temperature,
concentration of fuel and radicals respectively; $x$ and $t$ are
the dimensionless spatial coordinate and time respectively. In
(\ref{PDE}) the following non-dimensional parameters have been
introduced: $\beta= c_p E/QR$, $r = A_r/A_B$, $\tau_{A,B}  = \rho
c_p D_{A,B}/k$, where $c_p$ is the specific heat; $D_A$ and $D_B$
represent the diffusivity of fuel and radicals respectively, $A_r$
and $A_B$ are constants of recombination and chain branching
reactions respectively; $Q$ is the heat of the recombination
reaction; $E$ is the activation energy for chain branching
reaction; $R$ is the universal gas constant. Here parameter
$\beta$ is the dimensionless activation energy of the
chain-branching step which coincides with the corresponding
definition for the one step model \cite{SIAM}. Parameter $r$ is
the ratio of the characteristic time of the recombination and
branching steps and cannot be reproduced in one step approximation
of the flame kinetics.

Equations (\ref{PDE}) are considered subject to the boundary
conditions
\begin{equation}\label{BC}
\begin{array}{lllllllll}
u = 0,&& v = 1,&& w = 0 && \text{for}&& x \rightarrow \infty,\\
u_x = 0,&& v_x= 0,&& w = 0 && \text{for}&& x \rightarrow -\infty.
\end{array}
\end{equation}
On the right boundary we have cold ($u = 0$) and unburned state
($v=1$), where the fuel has not been consumed yet and no radicals
have been produced ($w=0$). We also take the ambient temperature
to be equal to zero. As noted in \cite{Weber} this is a convenient
way to circumvent the so-called ``cold-boundary problem'' and it
does not change the generic behaviour of the system. On the left
boundary ($x \rightarrow - \infty$) neither the temperature of the
mixture nor the concentration of fuel can be specified. We only
require that there is no reaction occurring so the solution
reaches a stationary point of (\ref{PDE}). Therefore the
derivatives of $u$, $v$ are zeros and $w= 0$ for $x \rightarrow-
\infty$.

\subsection{Travelling wave solution}\label{CTF}

We seek a solution to the problem (\ref{PDE}), (\ref{BC}) in the
form of a travelling wave $u(x,t) = u(\xi)$, $v(x,t) = v(\xi)$,
and $w(x,t) = w(\xi)$, where we have introduced $\xi = x - ct$, a
coordinate in the moving frame and $c$, the speed of the
travelling wave. Substituting the travelling wave solution into
the governing equations we obtain
\begin{equation}\label{ODE}
\begin{array}{l}
u_{\xi\xi} + cu_{\xi} + r w = 0,\\
\tau_A v_{\xi\xi} + c v_{\xi} -\beta vw e^{-1/u} = 0,\\
\tau_B w_{\xi\xi} + c w_{\xi} + \beta vwe^{-1/u} - r \beta w = 0,
\end{array}
\end{equation}
and boundary conditions
\begin{equation}\label{BC_o}
\begin{array}{lllllllll}
u = 0,&& v = 1,&& w = 0 && \text{for}&& \xi \rightarrow \infty,\\
u_{\xi} = 0,&& v_{\xi}= 0,&& w = 0 && \text{for}&& \xi \rightarrow
-\infty.
\end{array}
\end{equation}
Following \cite{Zeld,Simon} we consider the case when Lewis
numbers for the fuel and the radicals are equal to unity. Although
this assumption simplifies the problem significantly, it still
allows the investigation of the generic properties of the system
(\ref{ODE}) and (\ref{BC_o}).

In the case $\tau_A = \tau_B = 1$ equations (\ref{ODE}) possess an
integral $ S = \beta u + v + w$. Using $S$ and the first boundary
condition in (\ref{BC_o}), equations (\ref{ODE}) can be reduced to
a system of two second order equations for temperature and fuel
concentration
\begin{equation}\label{ODEred}
\begin{array}{l}
u_{\xi\xi} + cu_{\xi} + r(1-\beta u- v)  = 0,\\
v_{\xi\xi} + c v_{\xi} -\beta v(1-\beta u- v) e^{-1/u} = 0.
\end{array}
\end{equation}
On the right boundary we require that $u=0$ and $v = 1$, whereas
on the left boundary ($\xi \rightarrow -\infty$) we modify the
boundary conditions as follows
\begin{equation}\label{BCleft}
\begin{array}{ccc}
u = \beta^{-1}(1 - \sigma), &&v = \sigma,
\end{array}
\end{equation}
where $\sigma$ denotes the residual amount of fuel left behind the
wave and is unknown until a solution is obtained. We note here
that at first glance, system (\ref{ODEred}) looks very similar to
the equations describing the dynamics of the one-step adiabatic
reaction model considered in \cite{Weber}. However, in contrast to
the one-step adiabatic case, equations (\ref{ODEred}) do not have
an integral which enabled further simplification in \cite{Weber}.
Moreover, boundary conditions (\ref{BCleft}) suggest that there
can be some fuel left behind the reaction zone, which is
impossible in the case of a one-step adiabatic reaction model.

\subsection{Existence of the travelling wave solution}\label{Ex}

The basic properties of the system (\ref{ODEred}) can be found
from the stability analysis of the fixed points on $\xi
\rightarrow \pm \infty$. On the right hand side, as $\xi
\rightarrow \infty$, the fixed point coordinates (or asymptotic
values of $u$ and $v$) are given as $u=0$ and $v=1$. We will refer
to this stationary point as the end point. The linearization of
(\ref{ODEred}) around these values suggests that the end point is
a saddle-node for all physical parameter values with corresponding
eigenvalues given as $\mu_1 = 0$, $\mu_2 = -c$, $\mu_{3,4} =  (- c
\pm \sqrt{c^2 + 4r\beta})/2$.

\begin{figure}
\setlength{\epsfxsize}{70mm} \centerline{\epsfbox{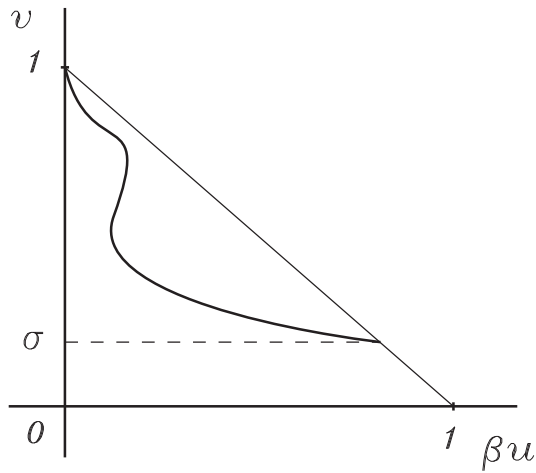}}
\caption{Schematic representation of the travelling wave solution
on $(\beta u, v)$ plane. }\label{uv}
\end{figure}

\begin{figure}\label{E_surf}
\setlength{\epsfxsize}{100mm} \centerline{\epsfbox{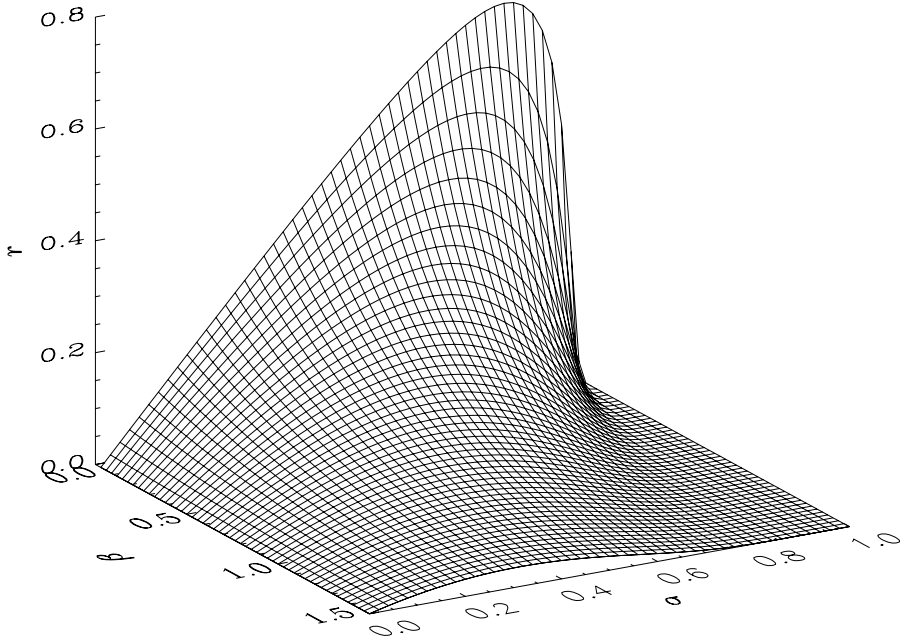}}
\caption{The extinction condition (\ref{ext_surf}) plotted as a
surface in the parameter space $(r,~\beta,~\sigma)$. }
\end{figure}

In the opposite limit $\xi \rightarrow -\infty$, the fixed point
(referred to as the starting point) has coordinates given by
(\ref{BCleft}), i.e. any point lying on a line $v = 1 - \beta u$
in $(u,~v)$ plane can be a fixed point (or correspondingly,
asymptotic values of temperature $u$ and fuel concentration $v$
belong to this line in the $(u,~v)$ plane as $\xi \rightarrow
-\infty$). On the $(\beta u,~v)$ plane a travelling wave solution
can be schematically represented as shown in figure \ref{uv}. The
linearization of the governing equations (\ref{ODEred}) around the
values (\ref{BCleft}) suggests that the starting point can be
either a saddle-node or a stable node depending on the parameter
values. Namely, the eigenvalues of the linearized problem can be
written in this case as: $\mu_1 = 0$, $\mu_2 = -c$, $\mu_{3,4} =
(-c \pm \sqrt{c^2 + 4\beta(r - \sigma e^{-\beta/(1-\sigma)})})/2$.
The starting point is a saddle-node for $r>\sigma
e^{-\beta/(1-\sigma)}$ and a stable node for $r<\sigma
e^{-\beta/(1-\sigma)}$. In the latter case it is impossible to
connect the starting point (as $\xi \rightarrow -\infty$) with the
end point (as $\xi \rightarrow \infty$) with a trajectory in the
phase space and therefore the solution does not exist. The
condition
\begin{equation}\label{ext_surf}
r = \sigma \exp{\displaystyle \frac{-\beta}{1-\sigma}}
\end{equation}
defines a surface in the $(r,~\beta,~\sigma)$ parameter space
which represents the boundary between the region, where the
solution exists from the region where it does not. If any of the
solution branches crosses this surface it has to exhibit
extinction. Therefore, we can also refer (\ref{ext_surf}) as the
extinction condition. In figure \ref{E_surf}, the boundary between
the parameter regions where the travelling solutions exist (above
the surface) or cease to exist (below the surface) is shown. The
physical meaning of this condition can also be explained if we
return to equation (\ref{PDE}) and neglect the diffusion terms.
Behind the travelling wave there has to be a stable state defined
by (\ref{BCleft}) and $w=0$, where no radicals can be produced if
some small variation of radical concentration is imposed. As seen
from (\ref{PDE}), the latter condition is satisfied if $r > \sigma
\exp{ \frac{-\beta}{1-\sigma}}$ and the right hand side of the
last equation in (\ref{PDE}) is not positive.

\begin{figure}
\setlength{\epsfxsize}{100mm}
\centerline{\epsfbox{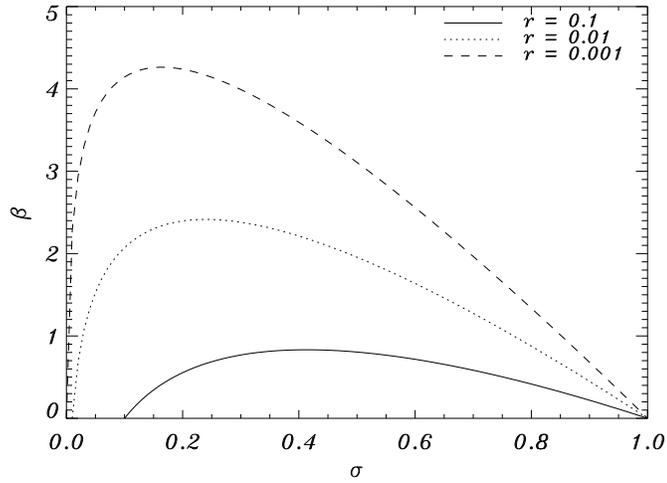}} \caption{The extinction
condition (\ref{ext_surf}) plotted as level curves for $r=0.1$,
$0.01$ and $0.001$. }\label{E_rconst}
\end{figure}

\begin{figure}
\setlength{\epsfxsize}{100mm} \centerline{\epsfbox{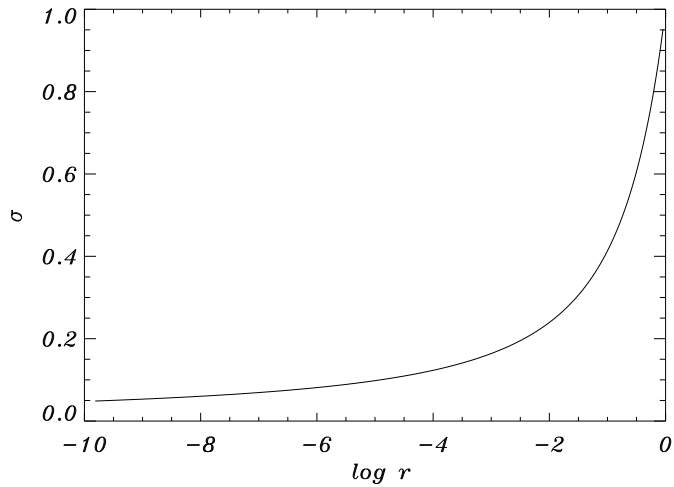}}
\caption{Dependence of maximal value of the residual amount of
fuel, $\sigma_m$, on $r$ for the extinction condition
(\ref{ext_surf}).}\label{sigma_m}
\end{figure}

\begin{figure}
\setlength{\epsfxsize}{100mm} \centerline{\epsfbox{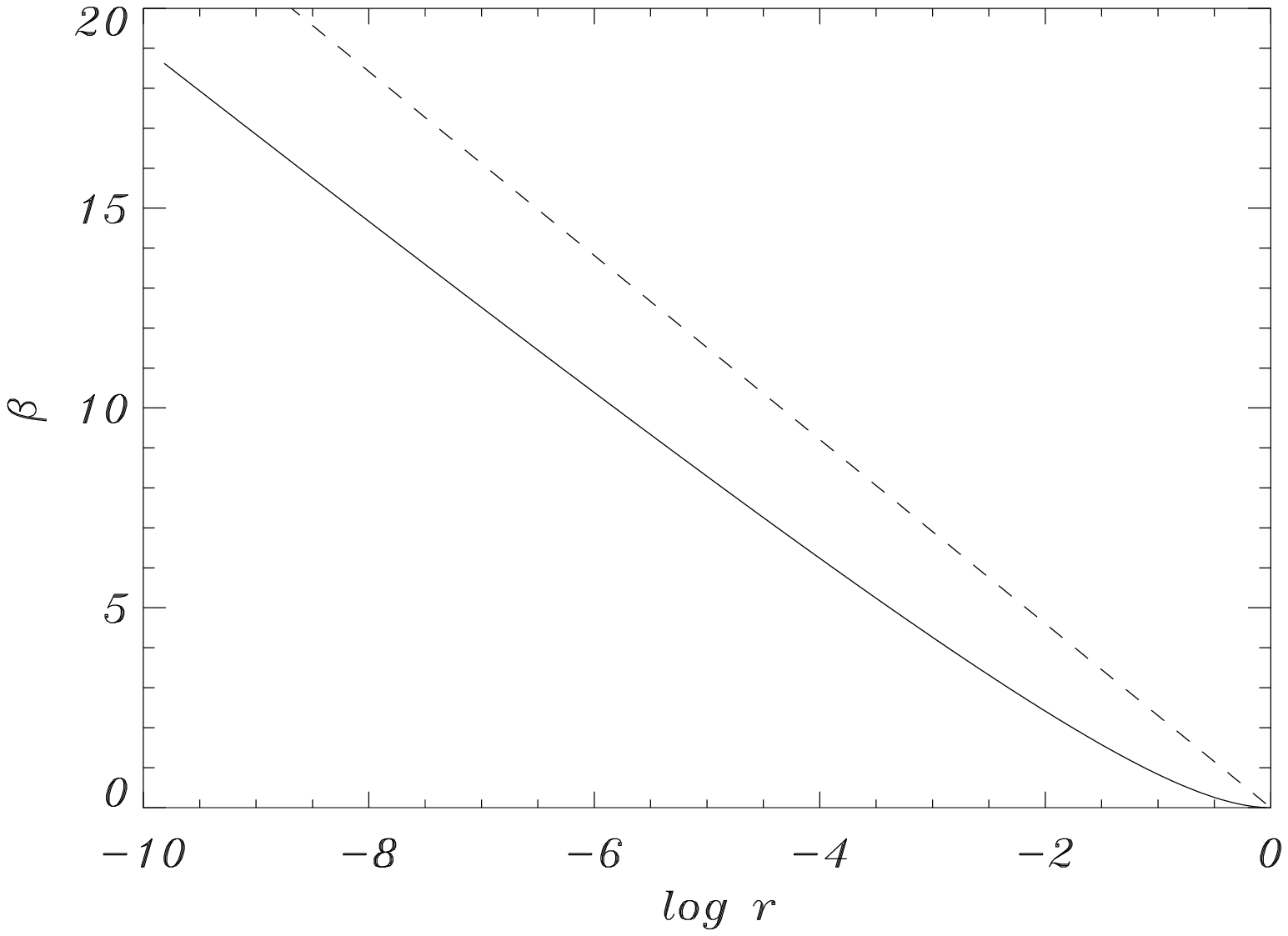}}
\caption{Dependence of $\beta_m$ on $r$ for the extinction
condition. The solid line represents the data plotted for
(\ref{ext_surf}), whereas the dashed line represents the results
obtained using the paper \cite{Dold04}.}\label{beta_mm}
\end{figure}

It is useful to consider several cross sections of
(\ref{ext_surf}) with the planes $r = constant$, since we will be
studying the properties of the solutions to (\ref{ODEred}) for
fixed values of $r$. The corresponding level curves are shown in
figure \ref{E_rconst} and are given as $\beta_e(\sigma) =
(1-\sigma)\ln(\sigma/r)$. It is seen that $\beta$ as a function of
$\sigma$ has a single maximum, $\beta_m$, for a fixed value of
$r$. We will denote the value of the residual amount of fuel for
which $\beta(\sigma)$ has maximum as $\sigma_m$, i.e. $\beta_m =
\beta(\sigma_m)$. This maximum is a solution to the $\partial
\beta_e/
\partial \sigma = 0$ equation which can be written as
$(1-\sigma_m)/\sigma_m - \ln{\sigma_m/r} = 0$ using equation
(\ref{ext_surf}). The solution of this equation is shown in figure
\ref{sigma_m} and corresponding value $\beta_m$ as a function of
$r$ is plotted in figure \ref{beta_mm}. It worthwhile noting that
according to \cite{Dold04} the condition for the extinction can be
expressed using our variables as $\beta \sim -\ln r$ and it is
shown in figure \ref{beta_mm} with a dashed line. The asymptotic
expression for the extinction condition was derived in the limit
$\beta\rightarrow\infty$, when $\sigma\rightarrow 0$. For moderate
values of the activation energy this dependence is given by
(\ref{ext_surf}) with the slope $1/(1-\sigma)$ times smaller.
Therefore the solid curve representing our results lies below the
dashed line corresponding to the asymptotic prediction of
\cite{Dold04}. For $r \rightarrow 1$ parameter $\sigma_m$ tends to
$1$ and the curve $\beta_m$ vs. $\log r$ becomes tangent to
abscissa axis. In the opposite limit $r\rightarrow 0$ parameter
$\beta_m$ tends to $\infty$ and $\sigma_m$ to zero, so that the
slope of both curves becomes equal.

It is interesting to compare the burning temperature which we
define as $u_b = (1-\sigma)/\beta$ with the so called crossover
temperature, $u_c$, a temperature at which the rate of branching
and recombination are equal given that $v = 1$ i.e. $u_c$: $r =
\exp( -1/u_c)$. Using this equation and relation
$r=\sigma_m\exp(-\beta_m/(1-\sigma_m))$ we derive $u_c^{-1} =
\beta_m/(1-\sigma_m) - \ln(\sigma_m)$. This yields
\begin{equation}
\frac{u_b}{u_c} =
\frac{1-\sigma}{\beta}\left(\frac{\beta_m}{1-\sigma_m} -
\ln{\sigma_m}\right) =
\frac{\beta_m}{\beta}\frac{1-\sigma}{1-\sigma_m}
\frac{1-\sigma_m(1+\ln\sigma_m)}{1-\sigma_m}.
\end{equation}
Since $\beta \leq \beta_m$, $\sigma \leq \sigma_m$ and
$\ln\sigma_m<0$ the first two factors in the right hand side of
the above equation are greater or equal to one and the last factor
is always greater than one. This indicates that the actual
temperature in the combustion wave is always above the crossover
temperature (even for the extinction condition). As discussed in
\cite{Dold03} the temperature of the flame has to be higher than
the crossover temperature so that the rate of the radicals
production can balance both their consumption due to chemical
reaction and to depletion of the radicals due to diffusion.


\section{Travelling front solution}

\begin{figure}
\setlength{\epsfxsize}{100mm} \centerline{\epsfbox{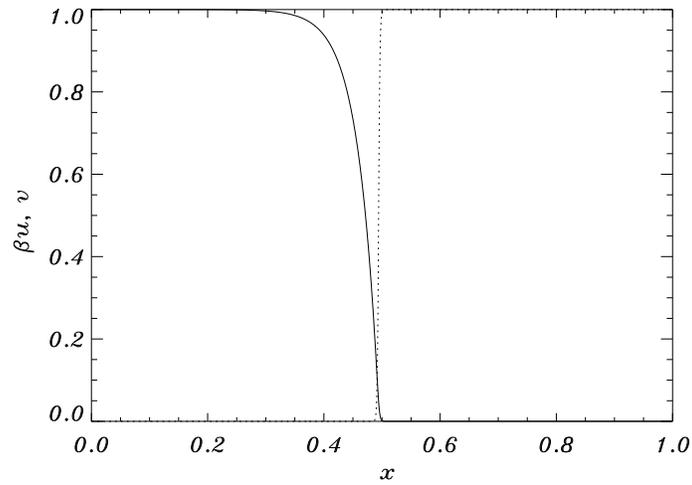}}
\caption{Classical travelling front solution profiles $u(\xi)$ and
$v(\xi)$ for $\beta = 0.054$, $r=0.01$. Solid line shows the
temperature profile and the dashed line represents the fuel
concentration.}\label{prof1}
\end{figure}

\begin{figure}
\setlength{\epsfxsize}{100mm} \centerline{\epsfbox{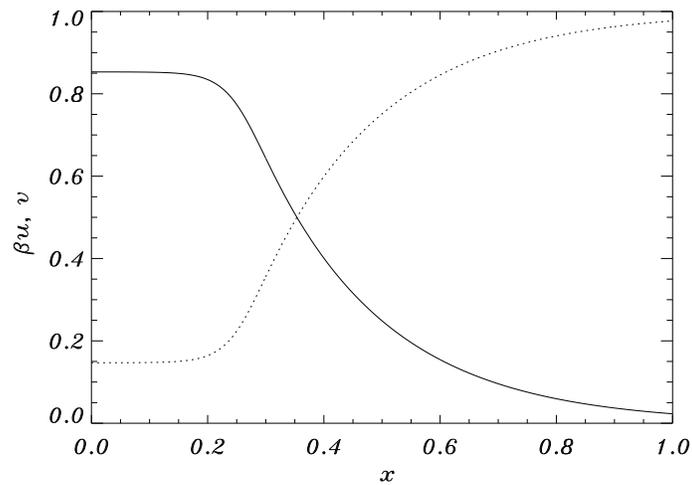}}
\caption{Classical travelling front solution profiles $u(\xi)$ and
$v(\xi)$ for $\beta = 2.364$, $r=0.01$. Solid line shows the
temperature profile and the dashed line represents the fuel
concentration.}\label{prof1a}
\end{figure}

In our previous paper \cite{JMC06} we have already shown that
equations (\ref{ODEred}) subject to the boundary condition
(\ref{BC_o}) exhibit travelling front solutions similar to those
found for the one-step reaction models, such that $u(\xi)$ and
$v(\xi)$ are monotonic and $u_{\xi}(\xi)$ and $v_{\xi}(\xi)$ are
bell-shaped functions of the spatial coordinate. Typical solution
profiles $u(\xi)$ and $v(\xi)$ are shown in figures \ref{prof1}
and \ref{prof1a}. The scaled coordinate $x$ is used instead of
$\xi$ so that $x \in [0,1]$. As seen from the figures the solution
profiles are sharper for smaller values of $\beta$ and flatter for
greater values of $\beta$. The residual amount of fuel $\sigma$
for fixed $r$ increases when $\beta$ increased.

\begin{figure}
\setlength{\epsfxsize}{100mm} \centerline{\epsfbox{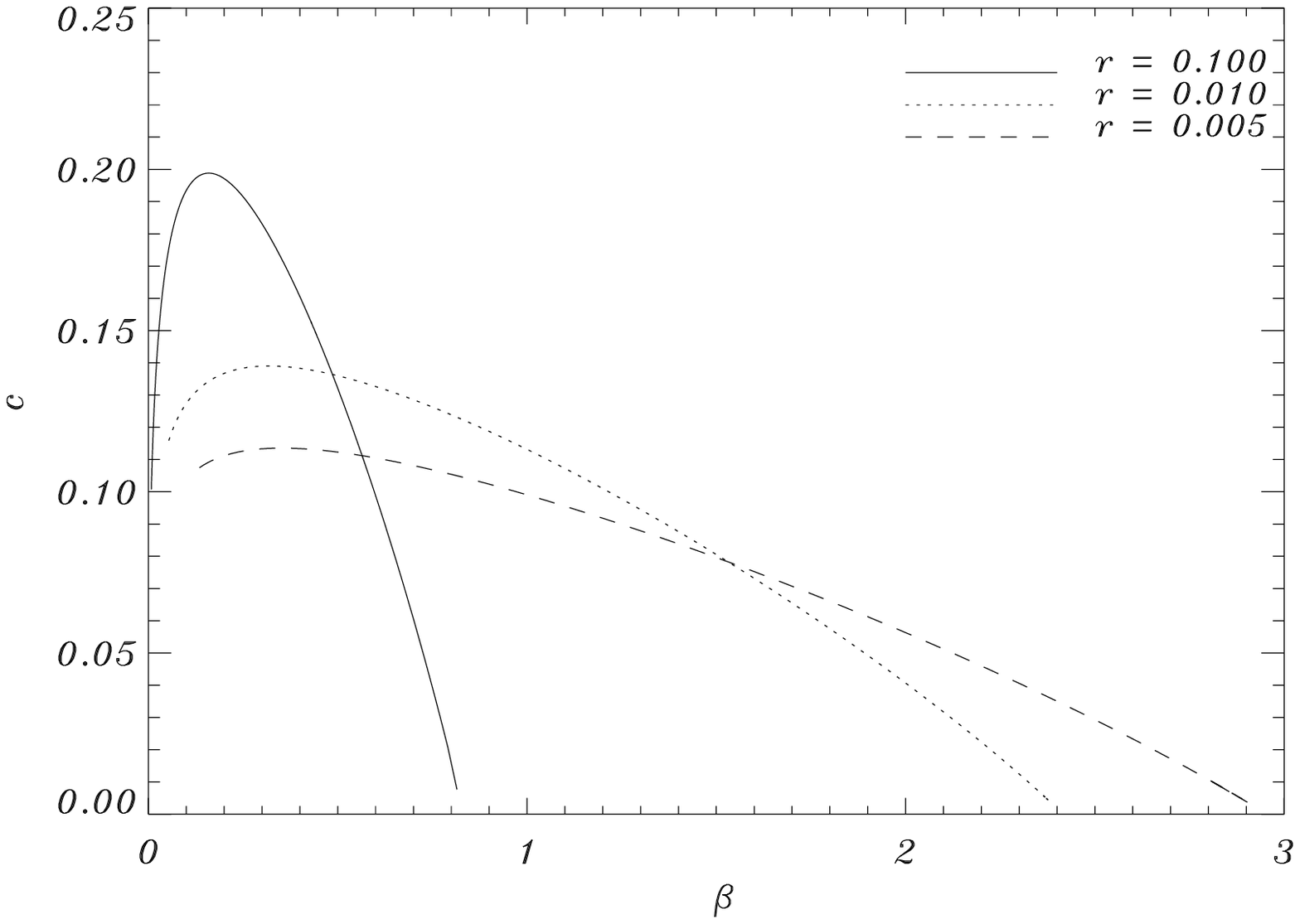}}
\caption{Speed of the front as a function of $\beta$ for various
values of $r$.}\label{c0}
\end{figure}

\begin{figure}
\setlength{\epsfxsize}{100mm}
\centerline{\epsfbox{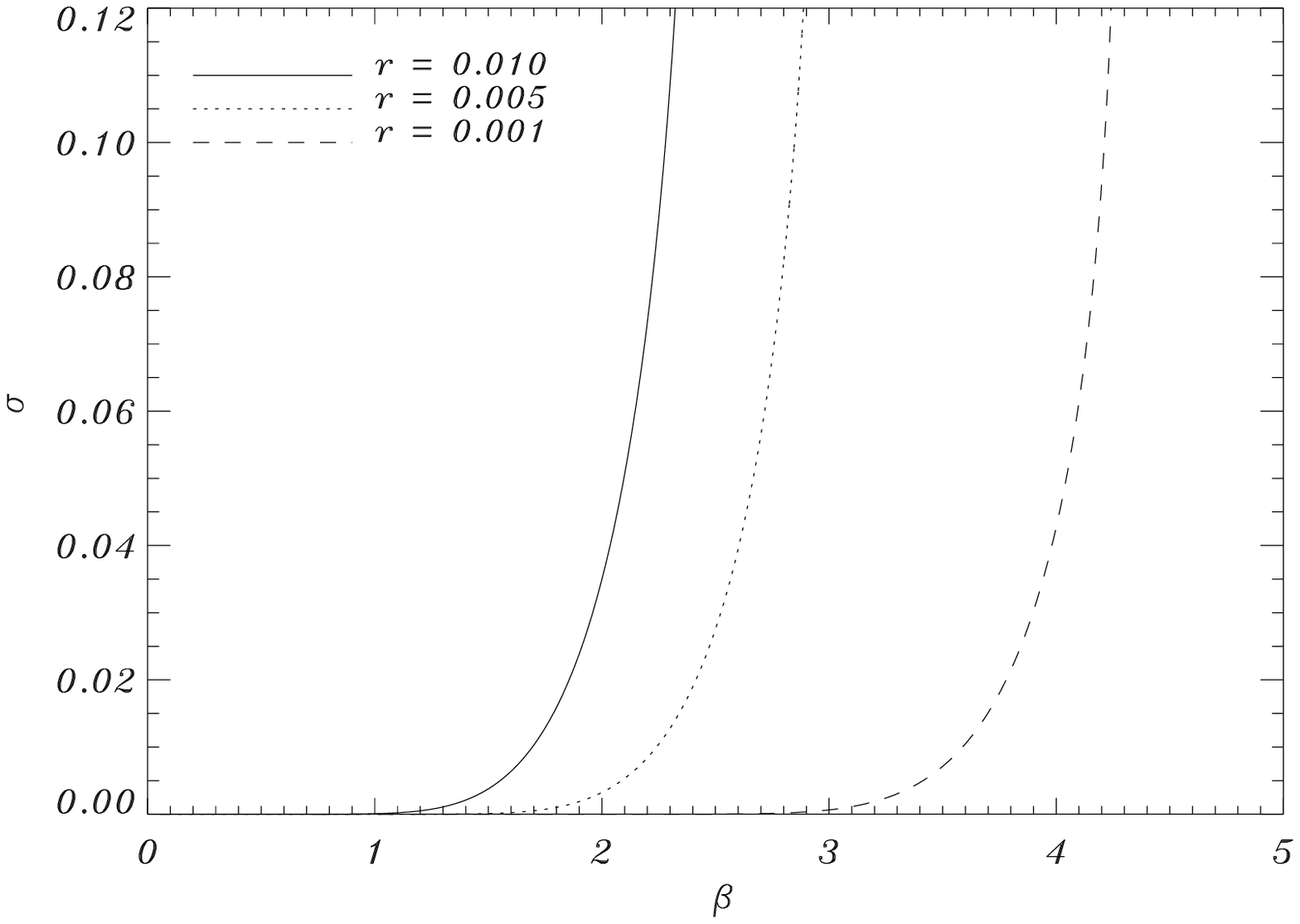}} \caption{Dependence of
$\sigma$ on $\beta$ for various values of $r$.}\label{sigma0}
\end{figure}

The properties of the classical travelling front solution are
summarized in figures \ref{c0} and \ref{sigma0}, where the speed
of the front $c$ and the residual amount of fuel $\sigma$ are
shown as functions of $\beta$ for several values of $r$.

As noted in \cite{JMC06}, at first glance the dependence of $c$
upon $\beta$ in figure \ref{c0} resembles the behaviour of the
speed of the front for the model with one-step reaction scheme,
which was studied in \cite{SIAM}. Namely, $c$ reaches the global
maximum for some value of $\beta$ and decays monotonically as we
increase (or decrease) $\beta$ from the value corresponding to the
maximum. However more detailed investigation shows that there is a
significant difference between the prediction of the one- and
two-step models. In particular, for the model with one-step
reaction mechanism the travelling front solution exists for any
value of $\beta$ and decays exponentially to zero as we increase
$\beta$. This is not the case for the model with two-step reaction
mechanism studied here.  As we increase $\beta$ (for a fixed $r$)
up to some critical value $\beta_c$, the speed of the front decays
rapidly and the travelling front solution ceases to exist for
$\beta> \beta_c$. Furthermore there is some residual amount of
fuel left behind the front in the case of the two-step model
unlike the one-step adiabatic model which has zero residual amount
of fuel. To some extent, the properties of the two-step adiabatic
model studied here resemble the properties of the nonadiabatic
one-step model investigated in \cite{ProA}, more than the
adiabatic one-step studied in \cite{SIAM}. This is expected since
the recombination step, which is an inhibitor of the chain
branching reaction, plays a similar role as the heat exchange with
the surrounding in the one-step nonadiabatic model. In both cases
there is a nonzero residual amount of fuel left behind the
reaction zone. The similarity between these two cases is also
strengthened by the likeness of the behaviour of the speed of the
front as a function of the parameter $\beta$. Namely, in both the
one-step nonadiabatic and the two-step adiabatic models, the
travelling front solution ceases to exist as we approach some
critical value of $\beta_c$ (in the combustion literature this
event is usually called extinction \cite{ProA}). However, the
route to extinction in these models appears to be different. In
the case of the one-step nonadiabatic model, for given parameter
values, there are either two solution branches with different
speeds or no solutions. The extinction occurs when the two
solution branches coalesce (this event is also known as a turning
point or a fold bifurcation). For the two-step reaction mechanism
the extinction occurs when the speed of the front drops down to
zero as we approach the critical parameter values.

It is evident that the route to extinction distinguishes this
model from its well known one-step counterparts. In the next
section we shall investigate the important phenomenon of
extinction in greater detail.

\subsection{Route to extinction}

\begin{figure}
\setlength{\epsfxsize}{100mm} \centerline{\epsfbox{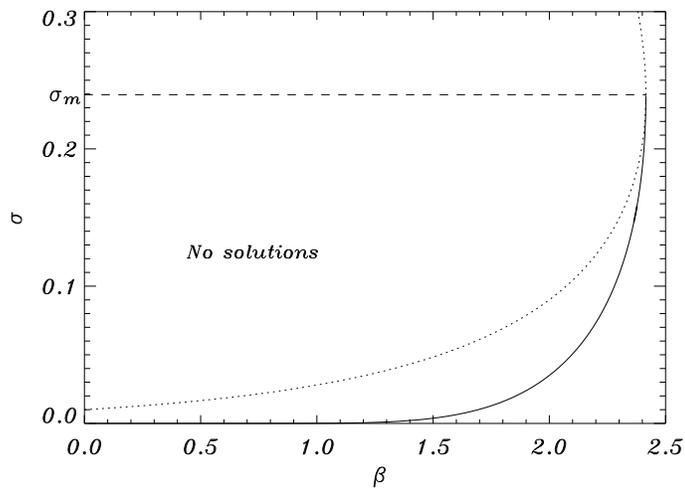}}
\caption{Parameter values $\sigma$ vs. $\beta$ for the travelling
front solution with $r=0.01$. Thick solid curve shows the
numerical results. Dotted curve represents the extinction curve
i.e. intersection of surface (\ref{ext_surf}) with plane $r=0.01$.
Dashed line marks the value $\sigma_m$ at which the extinction
curve reaches the maximum value of $\beta$.}\label{ext_rout}
\end{figure}

Previously we mentioned that extinction may occur when the
parameter values of the problem reach the boundary of the
existence of the solution (\ref{ext_surf}). Does this imply that
the the extinction of the classical travelling front solution
occurs when the parameter values of the problem approaches the
surface (\ref{ext_surf})? Extensive numerical analysis shows that
for fixed $r$ the travelling front solution exhibits extinction as
$\sigma$ and $\beta$ reach the values $\sigma_m$ and $\beta_m$
(the parameter values where the curve $\beta_e(\sigma)$, which is
an intersection of surface (\ref{ext_surf}) with plane $r =
const$, folds or the maximum value of $\beta$ is reached for fixed
$r$). This is illustrated in figure \ref{ext_rout} where the
dependence of $\sigma$ on $\beta$ is plotted for $r=0.01$ with a
thick solid curve. The critical parameter values for the
extinction are also shown on the same graph with a dotted curve.
As seen the travelling front solution ceases to exist as $\beta
\rightarrow \beta_m \approx 2.4153239$ and $\sigma \rightarrow
\sigma_m \approx 0.23947057$.

Next we investigate the behaviour of the travelling front solution
for parameter values close to the extinction point. Firstly, we
note that as we approach the extinction point along the travelling
solution branch, $\beta \rightarrow \beta_m$, $\sigma \rightarrow
\sigma_m$, $c \rightarrow 0$, $\sigma e^{-\beta/(1-\sigma)}
\rightarrow r$. In order to investigate the problem further we
split the solution profile into several zones according to the
physical processes dominating in these regions. The region in
front of the travelling wave, which is usually referred to as a
preheat zone, is governed by two processes: diffusion and
convection. For this reason it is also sometimes called the
diffusion-convection region. In this region the reaction terms are
negligible and can be omitted from the governing equations.
Without loss of generality let us assume that the preheat region
satisfies $\xi>\xi_b$. Since the travelling wave solution has a
translational invariance, $\xi_b$ can be chosen to be arbitrary.
For $\xi>\xi_b$ the front satisfies the following equation
\begin{equation}\label{preheatODE}
\begin{array}{l}
u_{\xi\xi} + cu_{\xi}  = 0,\\
v_{\xi\xi} + c v_{\xi} = 0,
\end{array}
\end{equation}
and boundary conditions
\begin{equation}\label{BC_preheat}
\begin{array}{lllllll}
u = 0,&& v = 1,&& \text{for}&& \xi \rightarrow \infty.\\
\end{array}
\end{equation}
A solution to (\ref{preheatODE}) subject to (\ref{BC_preheat}) can
be written explicitly as
\begin{equation}\label{pre_sol}
\begin{array}{lll}
u(\xi) = u_b e^{-c \xi},&& v(\xi) = 1 - (1 - v_b) e^{-c \xi},
\end{array}
\end{equation}
where $u_b$ and $v_b$ have to be found by matching the solution
with those of $\xi<\xi_b$. In the region $\xi<\xi_b$ the governing
physical processes which occur in this zone are reaction and
diffusion, and therefore is sometimes called the
reaction-diffusion zone. The convection process is negligible when
compared to other terms and can be omitted from the governing
equations. Re-scaling the temperature as $u \rightarrow \beta u$,
spatial coordinate as $z \rightarrow \sqrt{\beta r} \xi$ and
dropping the convection terms we can rewrite the governing
equations as
\begin{equation}\label{prodODE1}
\begin{array}{l}
u_{zz} + (1- u- v)  = 0,\\
\\v_{zz}  - \displaystyle
\frac{v}{r}(1- u- v) e^{-\beta/u} = 0,
\end{array}
\end{equation}
which has to be solved subject to the boundary conditions
\begin{equation}\label{BC_prod}
\begin{array}{lllll}
u = 1 - \sigma, &&v = \sigma, && z \rightarrow -\infty.
\end{array}
\end{equation}
In the opposite limit $z \rightarrow z_b$ we require that the
reaction terms vanish as $z \rightarrow z_b$, which implies
immediately that $u(z_b) + v(z_b) = 1$ (or $w=1-u-v \rightarrow
0$). Moreover, since the reaction terms are neglected in the
preheat region we assume that derivatives of $u(z) + v(z)$ also
vanish at $z =z_b$. The parameter $z_b$ is undefined and we shall
take it to be a sufficiently large number and replace the problem
on the half line $z\in [-\infty,z_b]$ by the problem on an
infinite line $z\in [-\infty,\infty]$ which is more convenient for
further investigation, noting that this expansion of the domain
does not impact on our results.

It is convenient to define new variables $u = 1 + \sigma u_1$ and
$v = \sigma v_1$ and use the substitution $r = \sigma_m
e^{-\beta_m/(1- \sigma_m)}$ to rewrite the governing equations as
\begin{equation}\label{prodODE2}
\begin{array}{l}
u_{1zz} -  u_1- v_1  = 0,\\
\\v_{1zz}  + \displaystyle
\frac{\sigma}{\sigma_m}v_1(u_1 + v_1) \exp[{\beta_m/(1 - \sigma_m)
- \beta/(1 - \sigma u_1)}] = 0,
\end{array}
\end{equation}
and boundary conditions as
\begin{equation}\label{BC_prod_2}
\begin{array}{lllllll}
u_1 = -1,&& v_1 = 1,&&\text{for}&& z \rightarrow -\infty,\\
u_1 + v_1 = 1,&& u_{1z} + v_{1z} = 0,&&\text{for}&& z \rightarrow
\infty.
\end{array}
\end{equation}
The solution to the problem (\ref{prodODE2}--\ref{BC_prod_2}) is
sought in the form of the series
\begin{equation}
\begin{array}{lllllll}\label{series}
u_1(z) = -1 + \displaystyle \sum_{i=1}^{\infty}{\delta^i
u_{1}^{(i)}(z)},&& v_1(z) = 1 + \displaystyle
\sum_{i=1}^{\infty}{\delta^i v_{1}^{(i)}(z)},
\end{array}
\end{equation}
with $\delta$ being a small parameter of the asymptotic expansion.
The physical interpretation of this parameter will be defined
later. As a next step we expand the exponential temperature
dependent factor in the second equation of (\ref{prodODE2}) in a
Taylor series about $u_1 = -1$:
\begin{equation}
\exp\left({\displaystyle -\frac{\beta}{1 - \sigma u_1}}\right)
\approx \exp \left({\displaystyle -\frac{\beta}{1 -
\sigma}}\right)\left(1 + \Lambda(u_1 + 1) + \Lambda\left(\Lambda -
2\frac{\sigma}{1-\sigma}\right) \frac{(u_1 + 1)^2}{2} +
...\right),
\end{equation}
where the notation $\Lambda = \sigma \beta/ (1-\sigma)^2$ has been
introduced and terms higher than quadratic in $u_1 + 1 \sim
O(\delta)$ have been neglected. Substituting this expansion into
(\ref{prodODE2}) yields
\begin{equation}\label{prodODE3}
\begin{array}{l}
u_{1zz} -  u_1- v_1  = 0,\\
\\v_{1zz}  + \displaystyle
K v_1(u_1 + v_1) \left(1 + \Lambda(u_1 + 1) + \Lambda
\left(\Lambda - 2\frac{\sigma}{1-\sigma}\right)\frac{(u_1 +
1)^2}{2}\right)  = 0,
\end{array}
\end{equation}
where parameter $K = \sigma/\sigma_m \exp[{\beta_m/(1 - \sigma_m)
- \beta/(1 - \sigma)}]$ has been introduced.

Since it is assumed that the parameters are chosen to be close to
the extinction values, we can therefore write $\sigma = \sigma_m +
\Delta \sigma$, where $\Delta \sigma \ll 1$. At the same time we
know that $\beta$ as a function of $\sigma$ is tangent to the
extinction curve $\beta_e(\sigma)$ at $\sigma = \sigma_m$.
Therefore it is reasonable to conclude that $\beta \approx \beta_m
+ \beta_2 (\Delta \sigma)^2 + ...$ since $\beta_e(\sigma)$ has a
maximum at $\sigma = \sigma_m$. The coefficient $\beta_2$ is equal
to $1/2(d^2 \beta/d\sigma^2)|_{\sigma = \sigma_m}$. We shall now
proceed to show the dependence of $K$ and $\Lambda$ on
$\Delta\sigma$. By differentiating (\ref{ext_surf}) with respect
to $\sigma$ for fixed $r$ it can be shown that $\sigma
\beta_e(\sigma)/(1 - \sigma)^2 \rightarrow 1$ as $\sigma
\rightarrow \sigma_m$. Therefore, it can be written that $\Lambda
\approx 1 +  \Lambda_1 \Delta \sigma + O((\Delta \sigma)^2) $,
where $\Lambda_1 = \beta_m(1+ \sigma_m)/(1 - \sigma_m)^3$. On the
other hand using the expansion for $\beta(\sigma)$ and $\Lambda$
it can be shown that $K \approx 1 - K_2 (\Delta \sigma)^2$, where
$K_2 = (1 + \sigma_m + 2\beta_2
\sigma_m^2)/(2\sigma_m^2(1+\sigma_m)^4)$. It is convenient to use
the small parameter defined by the equation $\delta = \Delta
\sigma K_2^{1/2}$ instead of $\Delta \sigma$. In this case we can
write the following asymptotic expansion for the parameter values
near the point of extinction
\begin{equation}\label{as_exp}
\begin{array}{l}
\sigma = \sigma_m + {K_2^{-1/2}} \delta,\\
\beta = \beta_m + \beta_2 {K_2^{-1}} \delta^2 + ...,\\
K = 1 - \delta^2+...,\\
\Lambda = 1 + \lambda_1 \delta + ...,\\
\end{array}
\end{equation}
where $\lambda_1 = \Lambda_1/ K_2^{-1/2}$.

Next we reduce the order of the differential equations
(\ref{prodODE3}) by redefining the new independent variable as $v$
instead of $z$ (here and later on, it is convenient to drop the
subscript '1' in (\ref{prodODE3})). This implies immediately that
we restrict the consideration only to the case of solutions with
monotonic dependence of $v$ on $z$. However, our travelling front
solution always satisfies this property, and therefore we do not
impose any restriction by replacing $z$ with $v$. Introducing $p
\equiv v_z$, and substituting $d/dz = p d/dv$ and $d^2/dz^2 =
pp_vd/dv + p^2d^2/dv^2$ into (\ref{prodODE3}) yields

\begin{equation}\label{prodODE4}
\begin{array}{l}
p^2 u_{vv} + pp_v u_{v} -  u- v  = 0,\\
\\p p_v  + \displaystyle K v(u + v) \left(1 + \Lambda(u + 1) + \Lambda
\left(\Lambda - 2\frac{\sigma}{1-\sigma}\right)\frac{(u +
1)^2}{2}\right)  = 0,
\end{array}
\end{equation}
subject to the boundary conditions
\begin{equation}\label{BC_prod_4}
\begin{array}{lllllll}
u = -1,&& p = 0,&&\text{for}&& v \rightarrow 1,\\
u + v = 0,&& u_{v} = -1 ,&&\text{for}&& v \rightarrow \infty.
\end{array}
\end{equation}
Earlier in this section we have noted that the travelling front
solution is sought in the form of an asymptotic series (see
equation \ref{series}) so we can expect that the solution profiles
change substantially on the length scale of the order of $\delta$
in $v$. On the other hand, it can be shown that close to $v=1$,
the solution to (\ref{prodODE4}) behaves as $u \sim -1 - K(v-1), u
+ v \sim \delta^2(v-1)$,  and $p \sim \delta (v-1)$ i.e. $u$,
$u+v$ and $p$ are of different order in terms of the small
parameter $\delta$. In order to take this into account we
introduce a new variable $s(v) \equiv u + v$ and seek the solution
in the form of
\begin{equation}\label{ps_expan}
\begin{array}{l}
s(v) = \delta^3 s_3(v) + \delta^4 s_4(v) + \delta^5 s_5(v)+ ...,\\
p(v) = \delta^2 p_2(v) + \delta^3 p_3(v) + \delta^4 p_4(v)+ ...
\end{array}
\end{equation}
We also re-scale the independent variable in (\ref{prodODE4}) as
$v = 1 + \delta y$, so that $y \in [0,\infty)$. Substituting
(\ref{ps_expan}) and the parameter expansion (\ref{as_exp}) into
(\ref{prodODE4}) and using $y$ as a new independent variable we
obtain a system of differential equations to each order of the
small parameter $\delta$. Combining these equations so as to leave
only the leading order terms $s_3$ and $p_2$ yields a closed
system of equations
\begin{equation}\label{P2S3}
\begin{array}{l}
p_2^2 s_3'' + p_2p_2's_{3} -  s_3(1 + \lambda_1 y + \kappa y^2)  = 0,\\
p_2 p_{2}'  + s_3  = 0,
\end{array}
\end{equation}
subject to the boundary conditions
\begin{equation}\label{BC_p2s3}
\begin{array}{lllllll}
p_2 = 0,&& s_3 = 0,&&\text{for}&& y \rightarrow 0,\\
p_2 \equiv V = const,&& s_3 = 0 ,&&\text{for}&& y \rightarrow
\infty.
\end{array}
\end{equation}

Here we have introduced $\kappa = (1+\sigma_m)/2(1-\sigma_m)$ and
the prime stands for $d/dy$. The leading order problem
(\ref{P2S3}--\ref{BC_p2s3}) includes two parameters $\lambda_1$
and $\kappa$. The latter is a fixed given number for each value of
$r$, whereas the former is an eigenvalue which has to be found
together with the unknown functions  $p_2(y)$ and $s_3(y)$. The
constant value $V$ is defined from the solution $p_2(y)$.

Once the problem (\ref{P2S3}--\ref{BC_p2s3}) has been solved for a
fixed $r$, the parameter values $\Lambda_1$, $K_2$ and $\beta_2$
can be found from their definitions above. The only unknown
parameter is the speed of the front, which can be found by
matching the solutions in the reaction-diffusion and preheat
zones. In order to undertake this matching we have to transform
the variables $p_2(y)$ and $s_3(y)$ back to the original variables
$u(\xi)$ and $v(\xi)$ which are used in equations (\ref{ODEred}).
To summarize the various sequences of variable changes we provide
the following list of transformations
\begin{equation}
\begin{array}{l}
u(\xi) \rightarrow^{(\ref{prodODE1})} \beta u(z)
\rightarrow^{(\ref{prodODE2})}\beta (1 - \sigma
u_1(z))\rightarrow^{(\ref{prodODE4})} \beta(1-\sigma u(v))
\\\rightarrow^{(\ref{P2S3})}\beta(1-\sigma (s(y) - 1 - \delta
y)),\\ \\
v(\xi) \rightarrow^{(\ref{prodODE1})} v(z)
\rightarrow^{(\ref{prodODE2})} \sigma
v_1(z))\rightarrow^{(\ref{prodODE4})} \sigma v
\rightarrow^{(\ref{P2S3})}\sigma (1 + \delta
y).\\
\end{array}
\end{equation}
The index above the arrows shows the equation number, where the
corresponding substitution has been made. In the leading order
$O(\delta^0)$ we have the matching condition: $u_b = 1-\sigma$ and
$v_b = \sigma$, where $u_b$ and $v_b$ are taken from
(\ref{pre_sol}). We also have to match the derivatives of the
solutions to the boundaries of the reaction-diffusion and preheat
regions. Solution of the preheat problem suggests that
$-du/d\xi|_{b} = dv/d\xi|_b = (1-\sigma) c$. On the other hand, it
follows from the boundary conditions (\ref{BC_p2s3}) that
\begin{equation}
\begin{array}{l}
\displaystyle \left. \frac{dv}{d\xi}\right|_b ~=~ \sigma_m
\sqrt{\beta_m r} \left. \frac {dv_1}{dz} \right|_b ~=~
\sigma_m \sqrt{\beta_m r} \lim_{v\rightarrow \infty}{p} ~=\\\\
\sigma_m \sqrt{\beta_m r} \delta^2 p_2(\infty)~ =~ \sigma_m
\sqrt{\beta_m r} V \delta^2.
\end{array}
\end{equation}
Therefore, we have the following expression for the speed of the
front
\begin{equation}\label{speed}
c = \sqrt{\beta_m r} \sigma_m V \delta^2/(1-\sigma_m),
\end{equation}
where terms of the order higher then $O(\delta^2)$ have been
dropped.

At a first glance, the asymptotic procedure of the speed
derivation is inconsistent. Indeed, the convection term in
equations (\ref{prodODE1}) which is proportional to $c \sim
\delta^2$ has been neglected. Nevertheless, from  equations
(\ref{prodODE3}) onwards the terms of the order of $\delta^2$ are
retained. This, however, does not lead to any inconsistencies
because if the convection terms are retained throughout the
analysis in the reaction-diffusion region, they still will not
effect the leading order equation (\ref{P2S3}) and will appear
only in the equations for the higher order of small parameter
$\delta$ (i.e. in the equations for $p_3$ and $s_4$ etc.).
Therefore in terms of the original parameters of the problem we
have
\begin{equation}\label{speed'n'beta}
\begin{array}{l}
\beta= \beta_m(r) - \beta_2(r)(\sigma - \sigma_m(r))^2, \\
c=\displaystyle \sqrt{\beta_m r}  \frac{\sigma_m}{1-\sigma_m} V(r)
K_2(r) (\sigma - \sigma_m(r))^2.
\end{array}
\end{equation}

\begin{figure}
\setlength{\epsfxsize}{100mm} \centerline{\epsfbox{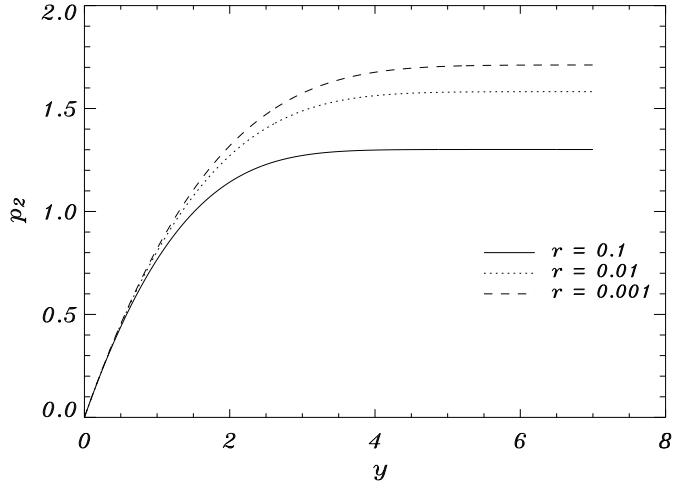}}
\caption{Solution profiles $p_2(y)$ for equations (\ref{P2S3}) for
various $r$ values.}\label{p2_prof}
\end{figure}

\begin{figure}
\setlength{\epsfxsize}{100mm} \centerline{\epsfbox{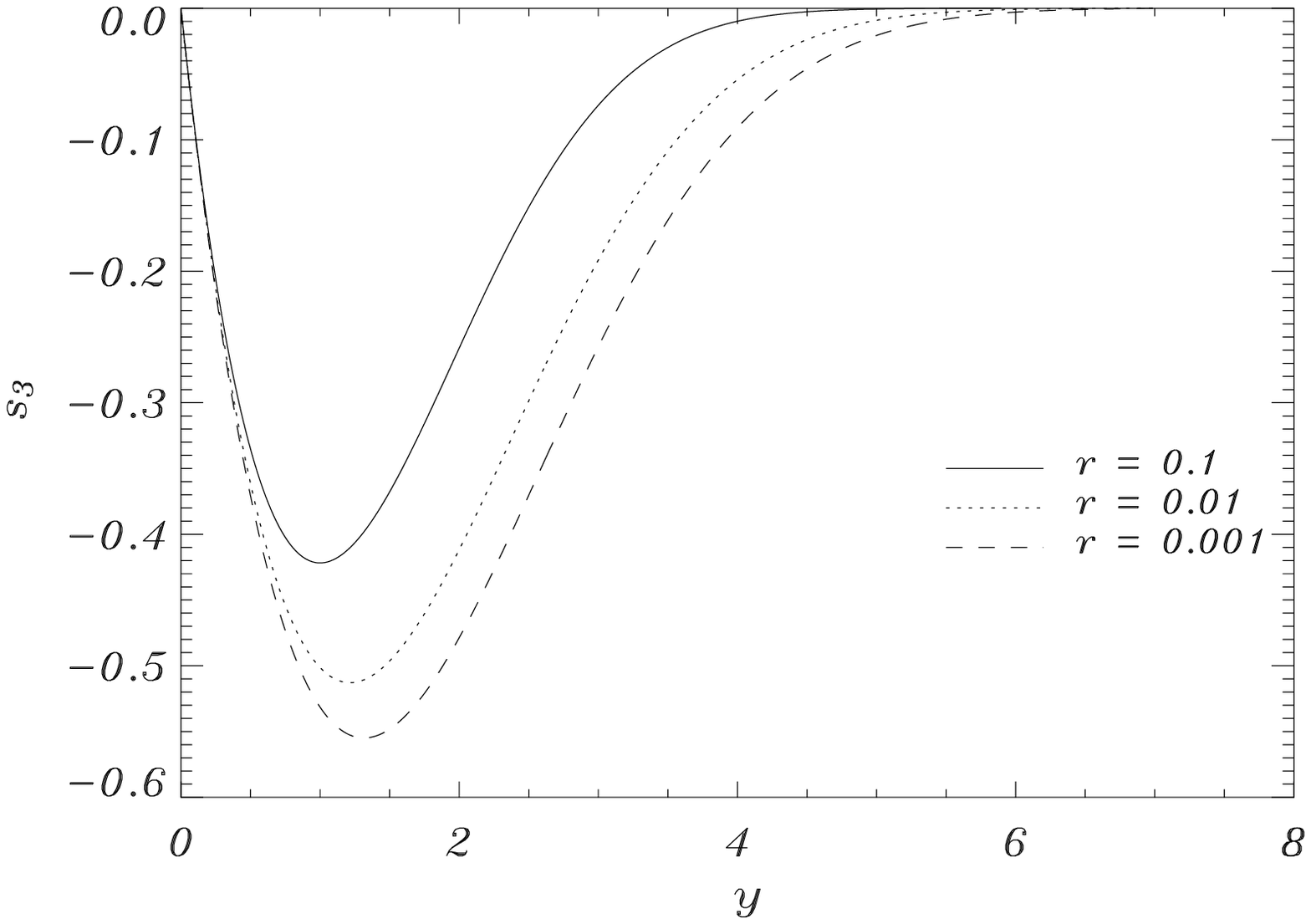}}
\caption{Solution profiles $s_3(y)$ for equations (\ref{P2S3}) for
various $r$ values.}\label{s3_prof}
\end{figure}

\begin{figure}
\setlength{\epsfxsize}{100mm} \centerline{\epsfbox{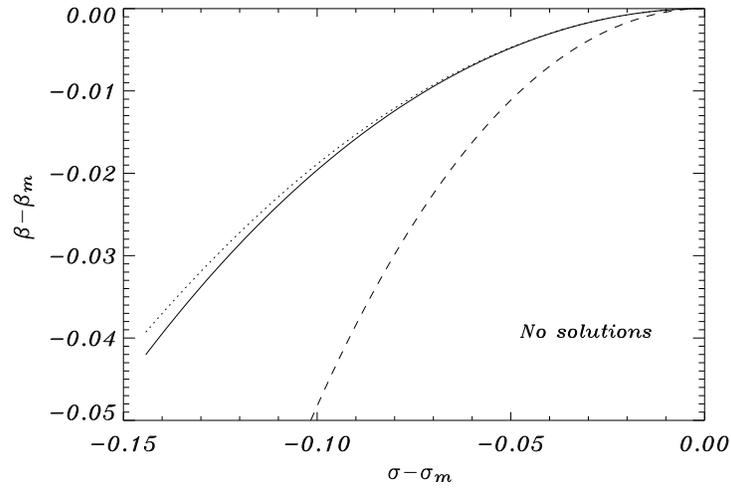}}
\caption{Dependence of $\beta$ on $\sigma$ for $r =0.1$. Solid
curve is plotted using the results of numerical integration of
(\ref{ODEred}). Dotted curve represents the prediction of
asymptotic formula (\ref{speed'n'beta}). Dashed curve shows the
extinction curve $\beta_e(\sigma)$ for $r=0.1$.}\label{B-S}
\end{figure}

\begin{figure}
\setlength{\epsfxsize}{100mm} \centerline{\epsfbox{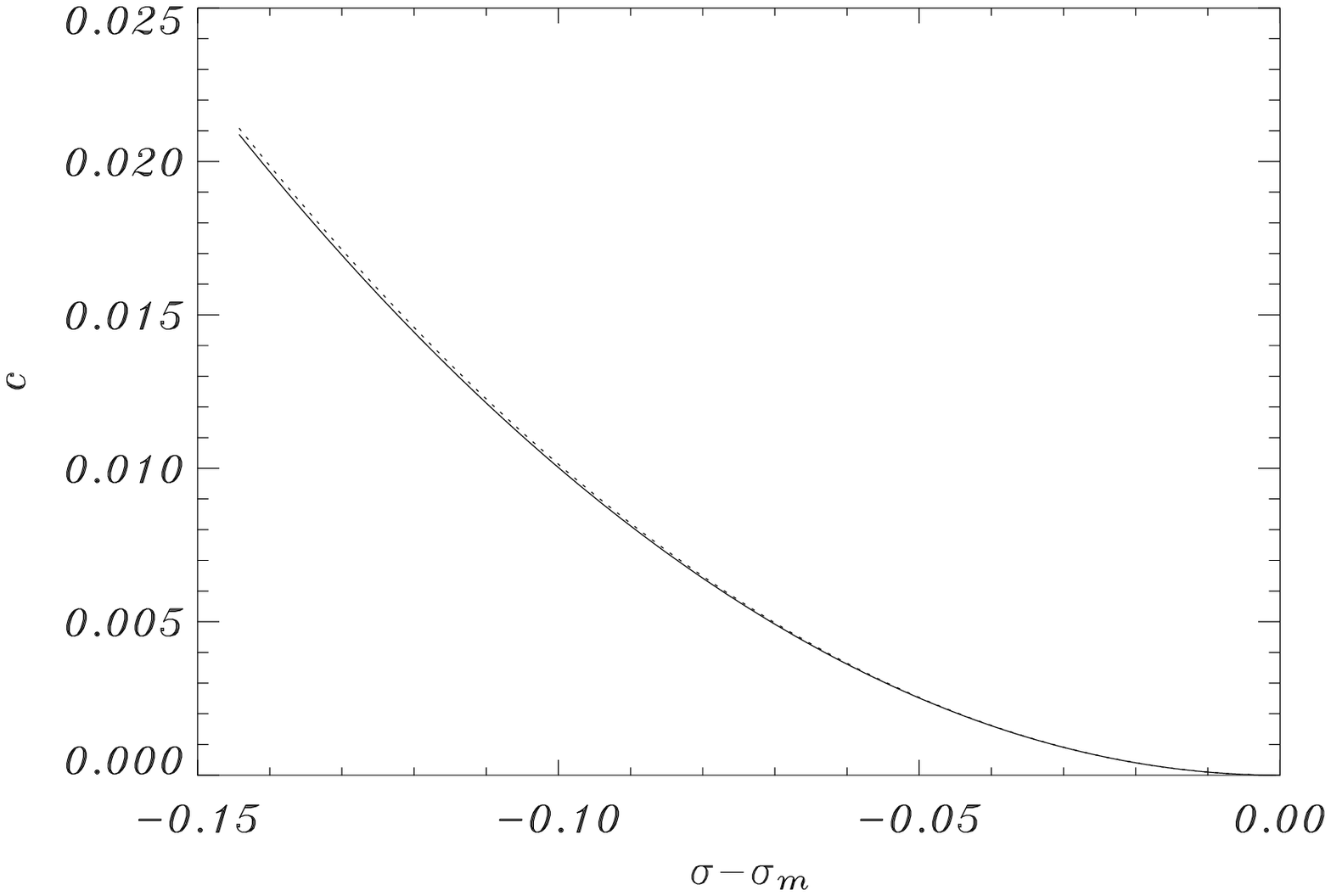}}
\caption{Dependence of $c$ on $\sigma$ for $r =0.1$. Solid curve
is plotted using the results of numerical integration of
(\ref{ODEred}). Dotted curve represents the prediction of
asymptotic formula (\ref{speed'n'beta}). }\label{C-S}
\end{figure}

In order to obtain estimates for $\beta(\sigma)$ and $c(\sigma)$
using equations (\ref{speed'n'beta}) we are required to solve the
system (\ref{P2S3}) numerically to obtain $\lambda_1$ and $V$. The
solution profiles $p_2(y)$ and $s_3(y)$ are shown in figures
\ref{p2_prof} and \ref{s3_prof}. For small values of $y$, both
$p_2$ and $s_3$ tends to zero linearly as $y$ and $-y$
respectively. In the opposite limit of large $y$ the value of
$s_3$ tends to zero. This causes the reaction terms in
(\ref{P2S3}) to vanish and $p_2$ to reach a constant value which
defines the speed of the front and the derivatives of the
temperature and the fuel concentration at the boundary between the
preheat and reaction-diffusion regions as discussed above. Once
the system (\ref{P2S3}) is solved for some $r$, the dependence of
the front speed $c$ and $\beta$ on the residual amount of fuel,
$\sigma$, can be found using (\ref{speed'n'beta}). In figures
\ref{B-S} and \ref{C-S}, $c$ and $\beta$ are shown as functions of
$\Delta \sigma = \sigma - \sigma_m$ for the values of $\sigma$
close to the extinction  value $\sigma_m$ and $r=0.1$. The
numerical results obtained by solving (\ref{ODEred}) are presented
on both figures with a solid line, whereas the dotted lines shows
the prediction of the asymptotic analysis (\ref{speed'n'beta}). In
figure \ref{B-S} we also plotted the extinction curve
$\beta_e(\sigma)$ with a dashed line. As seen from these figures
the correspondence between the numerical and asymptotic results is
excellent for $\sigma \rightarrow \sigma_m$. As $|\Delta \sigma|$
increases up to the value of $0.1$, the higher order terms in the
asymptotic expansion begin to play a significant role and the
discrepancy between the numerical and asymptotic results becomes
visible.

\begin{figure}
\setlength{\epsfxsize}{100mm} \centerline{\epsfbox{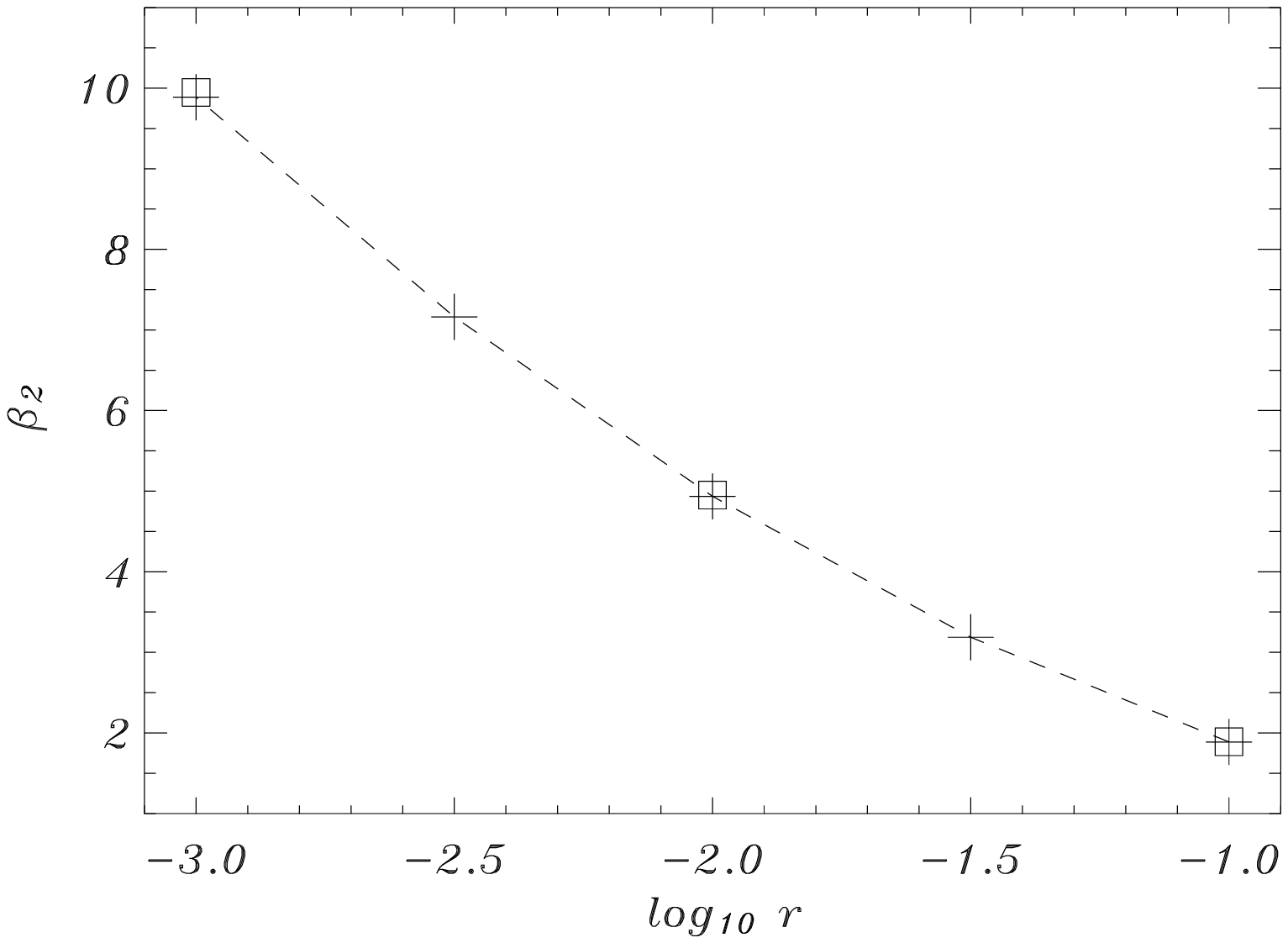}}
\caption{Dependence of $\beta_2$ on $r$. Crosses connected with
dashed line shows the results obtained with asymptotic formula
(\ref{speed'n'beta}). Squares represent $\beta_2$ estimated based
on the numerical solution of (\ref{ODEred}).}\label{b2}
\end{figure}

\begin{figure}
\setlength{\epsfxsize}{100mm} \centerline{\epsfbox{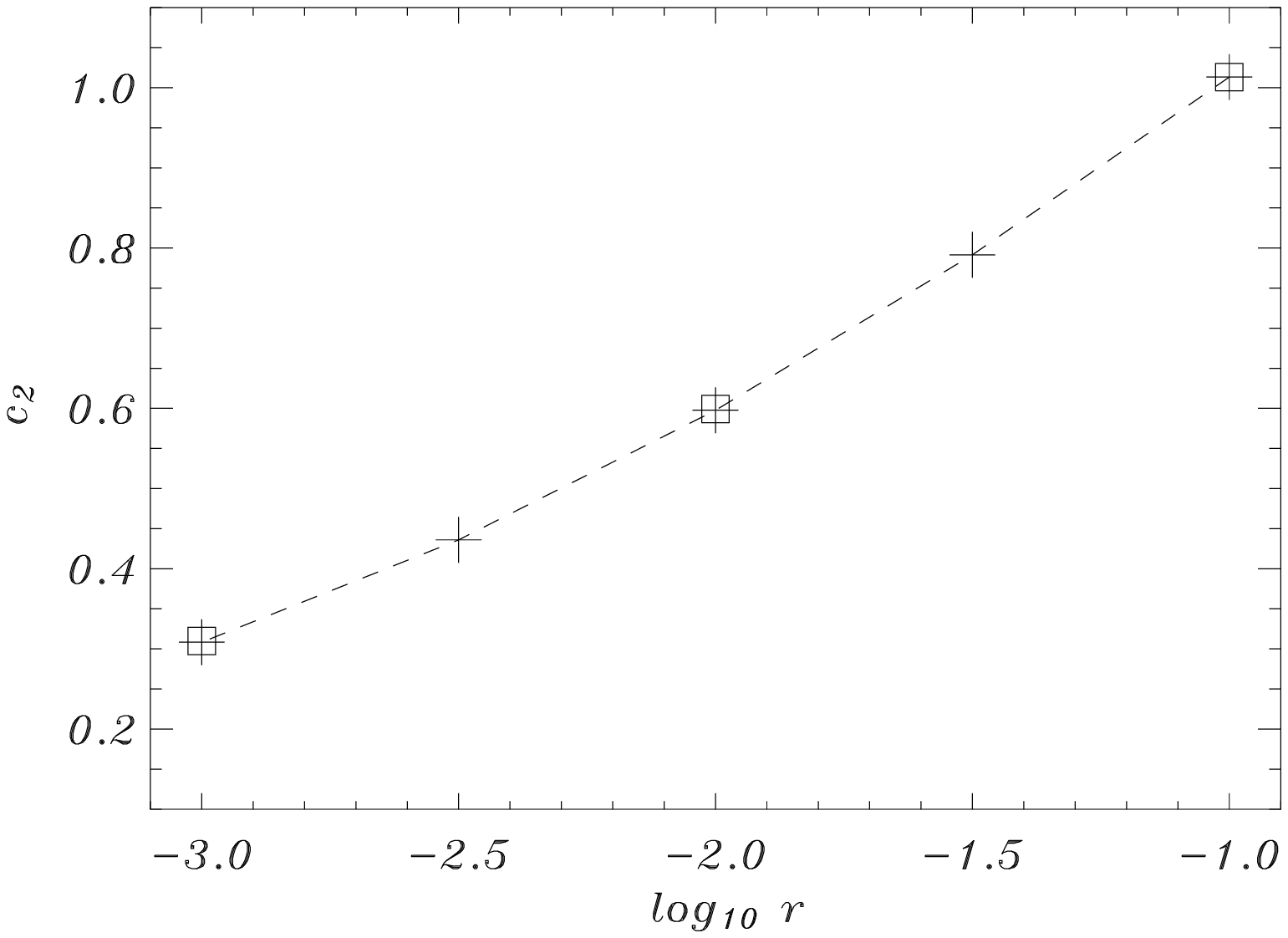}}
\caption{Dependence of $c_2$ on $r$. Crosses connected with dashed
line shows the results obtained with asymptotic formula
(\ref{speed'n'beta}). Squares represent $\beta_2$ estimated based
on numerical solution of (\ref{ODEred}).}\label{c2}
\end{figure}

We also compare the numerical and asymptotic results for different
values of $r$. In figure \ref{b2} and \ref{c2} we plot the
dependence of $\beta_2$ and $c_2$, which is $2d^2c/d\sigma^2$ as
$\sigma \rightarrow \sigma_m$, on $r$ respectively. Crosses
connected with dashed line show the results obtained analytically
by employing (\ref{speed'n'beta}) and squares denote the numerical
solution of (\ref{ODEred}) by calculating $2d^2\beta/d\sigma^2$ as
$\sigma \rightarrow \sigma_m$ in figure \ref{b2} and
$2d^2c/d\sigma^2$ as $\sigma \rightarrow \sigma_m$ in figure
\ref{c2}. The correspondence between the numerical and analytical
results is excellent for $r$ ranging over several orders of
magnitude. The discrepancy found in both cases was only in the
third significant digit.

\begin{figure}
\setlength{\epsfxsize}{100mm} \centerline{\epsfbox{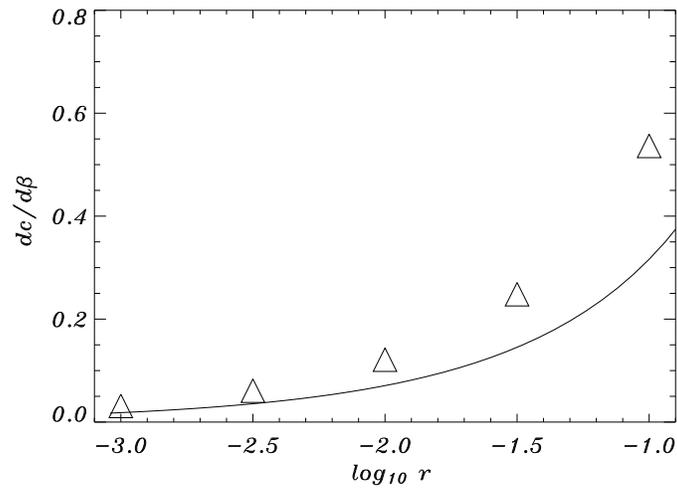}}
\caption{Dependence of $dc/d\beta$ on $r$. Triangles shows the
results obtained with asymptotic formula (\ref{speed'n'beta}).
Solid line represents the asymptotic results derived in
\cite{Dold04}.}\label{c_2_b}
\end{figure}

To conclude this subsection we shall summarize the main results
obtained here. We have investigated the behaviour of the
travelling front solution for parameter values close to the
extinction point. We determined that the extinction of the
travelling front solutions occurs when $\beta$ and $\sigma$ reach
the critical values $\beta_m$ and $\sigma_m$ corresponding to the
maximum condition for the function $\beta_e(\sigma)$ determined by
(\ref{ext_surf}) for a fixed $r$. The analysis of the solution
behaviour around the critical conditions $\beta_m$ and $\sigma_m$
shows that the solution branch on $\beta$ vs $\sigma$ follows a
quadratic law which is also the case for the dependence of the
speed of the front on $\sigma$ (see equation
(\ref{speed'n'beta})). The dependence of speed of the front vs.
$\beta$ is a linear function for $\beta$ close to $\beta_m$. Here
we note that the same type of behaviour for $c(\beta)$ is
predicted in \cite{Dold04}, although in  \cite{Dold04} $\sigma$ is
not considered and the authors employed different
nondimensionalization. In terms of the parameters used in this
paper the result of \cite{Dold04} can be written as $c \approx
\sqrt{r/\ln(1/r)} (\beta_m - \beta)$. In figure \ref{c_2_b} the
derivative of $c$ vs. $\beta$ at $\beta=\beta_m$ obtained by using
this formula (solid line) and calculated with asymptotic formula
(\ref{speed'n'beta}) (triangles) are compared. It is seen these
results are in good agreement. In the next subsection we proceed
to investigate the stability of the travelling front solution.

\subsection{Stability of the travelling front solution}

We investigate the stability of the travelling front solution
numerically using two methods: direct integration of the governing
PDEs and the Evans function method \cite{Evans72}. For both of
these methods it is convenient to rewrite (\ref{PDE}) using the
assumption of equal diffusivity of fuel, heat and radicals (i.e.
$\tau_A=\tau _B =1$) and the integral relation $S = \beta u + v +
w = 1$ (which is employed to reduce the governing equations
(\ref{ODE}) to (\ref{ODEred})) as
\begin{equation}\label{PDEred}
\begin{array}{l}
u_t = u_{xx}  + r(1-\beta u- v),\\
v_t = v_{xx}  -\beta v(1-\beta u- v) e^{-1/u},
\end{array}
\end{equation}
subject to the boundary conditions
\begin{equation}\label{BC_red}
\begin{array}{lllllllll}
u = 0,&& v = 1 &&\text{for}&& x \rightarrow \infty,\\
u_x = 0,&& v_x= 0 &&\text{for}&& x \rightarrow -\infty.
\end{array}
\end{equation}
In order to investigate the linear stability of the travelling
front solution $u(\xi)$, $v(\xi)$ using the Evans function method,
equations (\ref{PDEred}) are linearized around it and we seek the
solution of the form: $u(x,t) = u(\xi) +
\varepsilon\varphi(\xi,t)e^{\lambda t}$ and $v(x,t) = v(\xi) +
\varepsilon\chi(\xi)e^{\lambda t}$. Here $\varphi(\xi)$ and
$\chi(\xi)$ are linear perturbation terms,  $\varepsilon$ is a
small number, and $\lambda$ is the spectral parameter which is a
complex number. Substituting this solution into the governing
equations (\ref{PDEred}) and retaining terms linear in
$\varepsilon$ yields
\begin{equation}\label{L_stab}
\hat{L} \left(\begin{array}{l} \varphi\\
\chi
\end{array}\right) = \lambda\left(\begin{array}{l} \varphi\\
\chi
\end{array}\right),
\end{equation}
where
\begin{equation}\label{L}
\hat{L} =\left( \begin{array}{ccc} \partial^2_\xi + c\partial_\xi - r\beta  & - r\\
\displaystyle \frac{\beta v}{u^2}(\beta u^2 + \beta u+v-1)e^{-1/u}
& \partial^2_\xi + c\partial_\xi + \beta(2v+\beta u-1)e^{-1/u}\\
\end{array}\right).
\end{equation}
The stability of the travelling front is then defined from the
spectrum of {$\hat{L}$}. We can show that the essential spectrum
of this operator always lies in the left half plane by using the
approach outlined in \cite{Henry}. We consider the limiting
operators $\hat{L}^{\pm} = \lim_{\xi \rightarrow \pm
\infty}{\hat{L}}$. It can be shown that the essential spectrum of
$\hat{L}^{\pm}$ consists of algebraic curves
\begin{equation}\label{Ess}
\begin{array}{lllll}
\lambda_1(k) = - k^2 + i c k, && \lambda_2(k) = -  k^2 -
\beta\left(r-\sigma e^{-\beta/(1-\sigma)}\right) + i c
k,\\\lambda_3(k) = - k^2 - \beta r + i c k,
\end{array}
\end{equation}
where $k \in (-\infty, +\infty)$, which are located in the left
half plane and are symmetric about the real axis. The first curve
in (\ref{Ess}) includes the origin. Suppose that $K$ is the union
of the regions inside or on these curves, so that $\mathbf{C}
\backslash K$ contains the right half plane. Then according to
\cite{Henry}, the essential spectrum of $\hat{L}$ is contained in
$K$, and in particular includes curves (\ref{Ess}). Therefore the
essential spectrum of $\hat{L}$ lies in the left half plane
(including the origin) and the discrete spectrum is solely
responsible for the transition to instability. Namely, it can be
said that the travelling front is linearly unstable if, for some
fixed complex $\lambda$ with $Re(\lambda)>0$, there exists a
solution of (\ref{L_stab}) which decays exponentially as $\xi
\rightarrow \pm \infty$. We will refer to this $\lambda$ as an
eigenvalue and to the corresponding solution as an eigenmode.

We investigate the discrete spectrum of $\hat{L}$ using the Evans
function method. The Evans function $D(\lambda)$ is a function of
the spectral parameter $\lambda$ analytic in $\mathbf{C}
\backslash K$. It has the following properties: $D(\lambda) = 0$
if, and only if, $\lambda$ is an eigenvalue; the order of any zero
of the Evans function $D(\lambda)$ and the algebraic multiplicity
of the corresponding eigenvalue are equal. Therefore, the
stability analysis of the travelling front of (\ref{PDEred}) can
be reduced to the search for zeros of the Evans function located
in the right half plane. In our previous papers \cite{SIAM,ProA}
we have shown how to construct the Evans function and how to
calculate it numerically. In \cite{SIAM} it was demonstrated that
the number of zeroes $N$ located in the right half plane can be
calculated using the Nyquist plot technique. Obviously, if $N=0$
then the travelling front is stable and if $N>0$ then the
travelling front is unstable. We have calculated $N$ along the
travelling front solution branch of (\ref{PDEred}) for various
values of $r = 0.1$, $0.01$, and $0.001$. Parameter $\beta$ has
been taken for a range from $\beta < 0.01$ up to $\beta_m$ for
each value of $r$. In all cases we have found that $N=0$, that is
the travelling front solution is always stable. The same result
was obtained in \cite{Dold04} in the limit of high activation
energy.

\begin{figure}
\setlength{\epsfxsize}{100mm} \centerline{\epsfbox{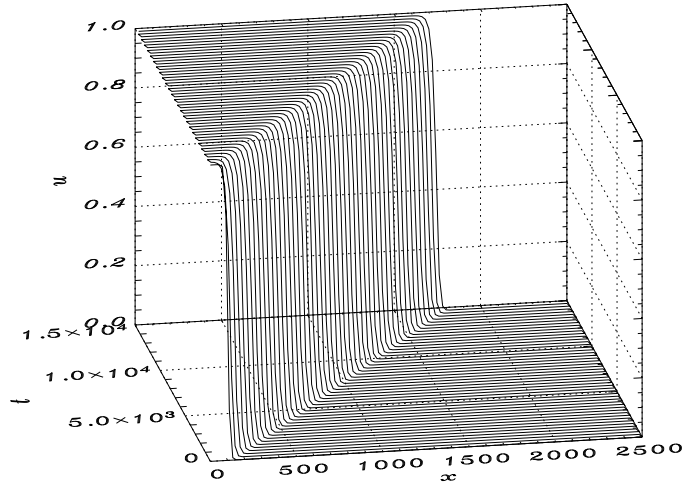}}
\caption{Temperature profile $u(x,t)$ obtained by numerical
integration of (\ref{PDEred}) for $\beta =1$ and
$r=0.01$.}\label{u}
\end{figure}

\begin{figure}
\setlength{\epsfxsize}{100mm} \centerline{\epsfbox{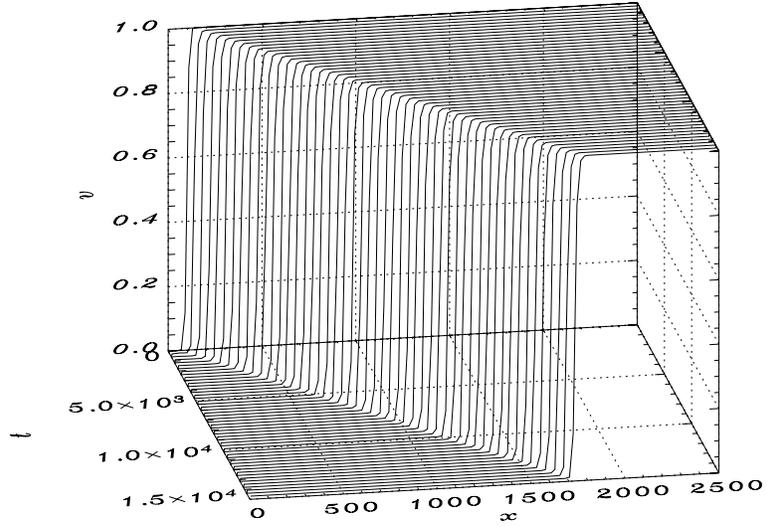}}
\caption{Fuel concentration profile $v(x,t)$ obtained by numerical
integration of (\ref{PDEred}) for $\beta =1$ and
$r=0.01$.}\label{v}
\end{figure}

We also investigate the stability of the travelling front solution
by direct integration of the governing PDEs (\ref{PDEred}) using
the time and space adaptive finite element package \cite{FlexPDE}.
We take the initial solution profiles in the form consistent with
the boundary conditions (\ref{BC_red}) as
\begin{equation}
\begin{array}{l}
u(x,0) = (2\beta)^{-1}(1-\sigma)(1 - \tanh[l(x-x_0)]), \\
v(x,0)= (1 + \sigma + (1 - \sigma) \tanh[l(x-x_0)])/2,
\end{array}
\end{equation}
where $l$ and $x_0$ are chosen in such a way so that $u(x,0)$ and
$v(x,0)$ fit the travelling front profiles $u(\xi)$ and $v(\xi)$
of (\ref{ODEred}). In figures \ref{u} and \ref{v} typical solution
profiles $u(x,t)$ and $v(x,t)$ of (\ref{ODEred}) are shown. The
initial conditions converge quickly to a travelling front solution
which propagate with constant speed. Note that the time axis in
figure \ref{v} is inverted in order to give a proper perspective.

\begin{figure}
\setlength{\epsfxsize}{100mm} \centerline{\epsfbox{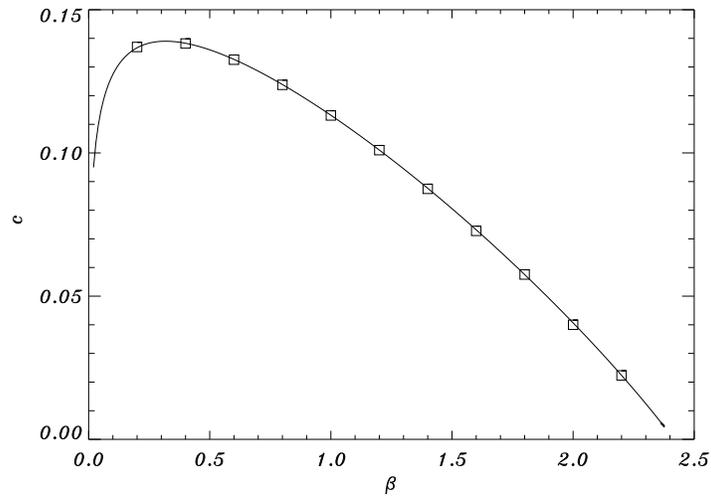}}
\caption{Speed of the travelling front solution as a function of
$\beta$ for $r=0.01$. The solid curve represents the results of
integrating the system of ODEs (\ref{ODEred}), whereas the squares
correspond to the results obtained from integrating the governing
PDEs (\ref{PDEred}) integration.}\label{c_pde}
\end{figure}

Solutions to (\ref{PDEred}) for different values of $\beta$ and
$r=0.01$ are also calculated. The travelling front is found to be
stable for all trial parameter values. In figure \ref{c_pde} we
plot the speed of the travelling front as function of $\beta$ for
$r = 0.01$. The numerical results obtained from direct integration
of the system of ODEs (\ref{ODEred}) (solid line) and PDEs
(squares) are represented on the same graph. The correspondence
between the prediction of both approaches is excellent. The
discrepancy was found only in the fourth significant digit.

\begin{figure}
\setlength{\epsfxsize}{100mm} \centerline{\epsfbox{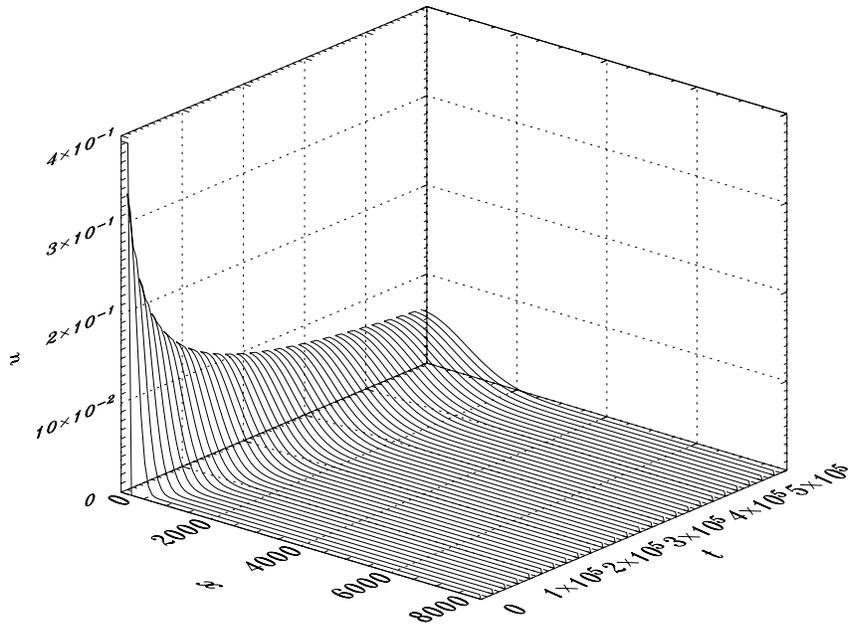}}
\caption{Temperature profile $u(x,t)$ obtained by numerical
integration of (\ref{PDEred}) for $\beta =2.5$ and
$r=0.01$.}\label{u1}
\end{figure}

\begin{figure}
\setlength{\epsfxsize}{100mm} \centerline{\epsfbox{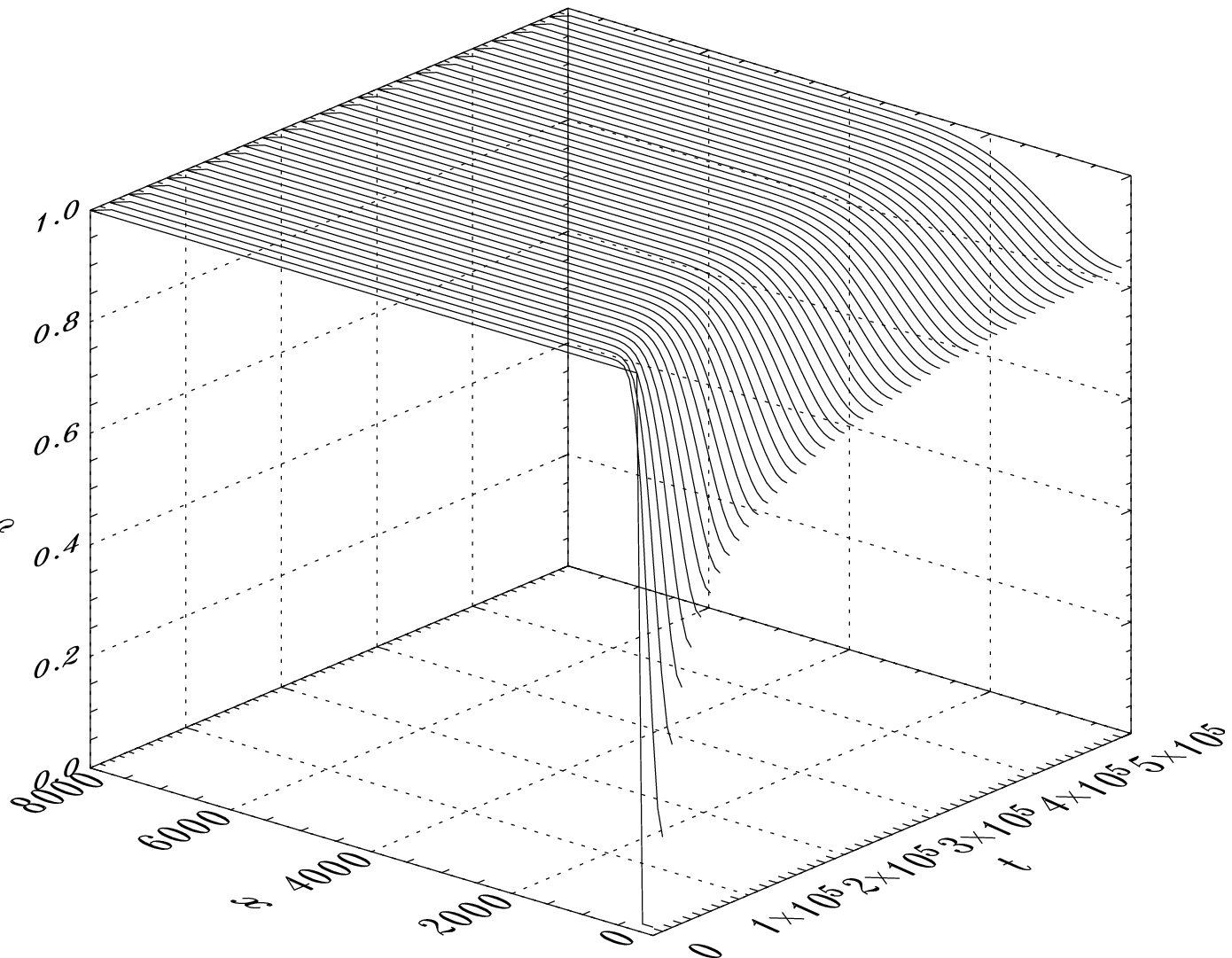}}
\caption{Fuel concentration profile $v(x,t)$ obtained by numerical
integration of (\ref{PDEred}) for $\beta =2.5$ and
$r=0.01$.}\label{v1}
\end{figure}

\begin{figure}
\setlength{\epsfxsize}{100mm} \centerline{\epsfbox{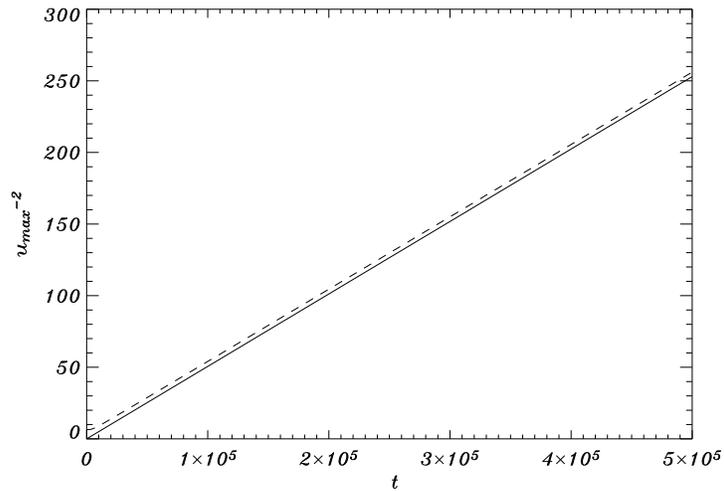}}
\caption{Decay of the maximum temperature $u(0,t)$ of
(\ref{PDEred}) solution with time for $r=0.01$ and $\beta = 2.5.$
Dashed line represent numerical results and solid line shows the
analytical estimate of decay rate.}\label{diffusion}
\end{figure}

It is interesting to study the behaviour of the solution of the
governing PDE (\ref{PDEred}) for various values of $\beta$ above
$\beta_m$ in order to investigate if the flame can exist beyond
the critical extinction condition. We integrate equations
(\ref{PDEred}) for $r=0.01$ and $\beta = 2.5 > \beta_m$. For all
trial initial conditions of the form discussed above, the solution
decays to the equilibrium state $u(x,t)=0$ and $v(x,t)=1$ with
typical solution profiles shown in figures \ref{u1} and \ref{v1}.
The maximum temperature $u(0,t)$ drops rapidly in time to zero and
subsequently the reaction ceases and the reaction terms vanish in
(\ref{PDEred}) and the system exhibits almost linear diffusion to
the equilibrium state. This is clearly seen in figure
\ref{diffusion} which shows the dependence of the squared inverse
maximum temperature upon time. The numerical results obtained with
the integration of the governing equations (\ref{PDEred}) (dashed
line) are compared with the analytical prediction (solid line),
which was found by considering equations (\ref{PDEred}) without
any reaction terms (i.e. we consider only the diffusion effects)
with initial conditions as described above. In this case it can be
shown that in the limit $t \rightarrow \infty$ the asymptotic
behavior of the maximum temperature satisfies $u(0,t)^{-2} = \pi
\beta^2 t/x_0^2(1-\sigma)^2$, which is shown with a solid line on
the figure \ref{diffusion}. The numerical results are shifted from
the asymptotic prediction, especially near the starting point
$t=0$. However for larger values of $t$ the maximum temperature
follows the diffusion law derived from the linear diffusion
equations.

\section{Conclusions and discussions}

In this paper we have studied combustion wave propagation in an
adiabatic model with two-step chain branching reaction mechanism.
The current work extends our previous preliminary investigation of
chain branching flames \cite{JMC06}. In contrast to \cite{Dold04}
we are not using the activation energy asymptotic approach and
consider the model for general values of activation energy. We
have also used different non dimensional parameters which enabled
a more convenient comparison between the properties of one- and
two-step models.

It is demonstrated that the model exhibits travelling combustion
front solutions. The properties of these solutions differ form the
properties of one-step models. Combustion waves are found to exist
in certain regions of the parameter space and are characterized by
non zero residual amount of reactant, $\sigma$, left behind the
travelling wave. This is not possible in one-step adiabatic flame
models as all the fuel is used up in such models. To some extend
this characteristic of the two-step adiabatic model is similar to
the properties of premixed combustion waves in nonadiabatic
one-step models, which can also exist in a certain region of
parameter space and exhibit extinction as the boundaries of this
region are reached. However the behaviour of the travelling
combustion waves near the extinction condition are completely
different in these two types of models. In the one step
nonadiabatic model the point of extinction corresponds to the
turning point of the dependence of flame velocity on parameters
(which is a two-valued function), therefore the speed of the wave
is always positive. In our two-step model studied here the flame
velocity is a single valued function and the velocity drops down
to zero as the extinction condition is reached. It is remarkable
that at the extinction condition there exist a flame characterized
by stationary distribution of temperature, reactant and radicals
i.e. a standing combustion wave.

The boundaries of existence for combustion front are determined.
We show that for a fixed value of $r$ (dimensionless recombination
rate) travelling wave solution branch occupy a certain region in
the parameter space which lies outside the region bounded by the
curve $\beta_e(\sigma)$ which has a maximum $\beta_m$ at
$\sigma_m$ so that $\beta$ remains less than or equal to $\beta_m$
and a residual amount of fuel less than $\sigma_m$. The maximum
values $\beta_m$ and $\sigma_m$ are achieved at the point of
extinction.

The properties of the travelling front solution are studied both
analytically using the matched asymptotic analysis and numerically
via the integration of the governing PDEs (\ref{PDEred}) and the
ODE formulation of the travelling wave solution (\ref{ODEred}).
The matched asymptotic analysis allowed us to obtain the
properties of the travelling front solution near the point of
extinction. The asymptotic expression for the speed of the front
is obtained as well as the location of the solution branch in the
parameter space. In particular, it is shown that for a fixed value
of $r$ the speed of the front and $\beta$ are quadratic functions
of $\sigma - \sigma_m$ near the point of extinction. The results
of the asymptotic analysis are then compared with the results
obtained by the integration of the governing ODEs for the
travelling wave solutions by the relaxation method as described in
\cite{SIAM,ProA}. It is demonstrated that the results obtained
from these two approaches are in excellent agreement. The speed of
the front and the location of solution branch in the parameter
space are studied in detail for a broad range of parameter values
by means of numerical integration of the ODEs. The results of this
investigation are then compared with the direct numerical
integration of the governing PDEs and are shown to be in an
excellent agreement. The stability of the travelling front
solution is investigated numerically by employing the Evans
function method. It is shown that the classical travelling front
solution is stable for all parameter values. These results are
also confirmed by direct numerical integration of the governing
PDEs. We also compare the results of our paper with the results
obtained in \cite{Dold04} in the limit of high activation energy
and show that they qualitatively agree.

In this paper we considered the case when the diffusivity of fuel
and radicals are equal: $\tau =\tau_A = \tau_B$. Moreover, for the
sake of simplicity it is assumed that these diffusivity are both
equal to the heat diffusivity: $\tau = 1$. Using these assumptions
we found that the classical travelling solution is stable for the
whole range of parameters $r$, $\beta$, and $\sigma$ where the
solution exists. Similar situation was observed for the one-step
nonadiabatic flames. In \cite{ProA} it was shown that for $\tau=1$
one of the travelling front solution branches was stable for all
parameter values where the solutions exist. However, for different
values of $\tau<1$ the situation changes due to the presence of
the Bogdanov-Takens bifurcation. Therefore, one of the most
important issues for further investigation is to understand how
the stability of the solutions changes in the chain-branching
model for different values of $\tau$.

\section{Acknowledgments}
The authors are thankful to R.O. Weber for his helpful
discussions. VVG would like to thank the University of New South
Wales for the grant PSO7264 which enabled him to visit UNSW at
ADFA, where the majority of this work was completed and
acknowledge the financial support from the Russian Foundation for
Basic Research: grants 05-01-00339, 05-02-16518, 06-02-17076 and
Russian Academy of Sciences: program "Problems in Radiophysics".

\end{document}